\documentclass[prxlongbibliography,twocolumn,showpacs,nofootinbib,superscriptaddress,notitlepage]{revtex4-1}
\usepackage{graphicx}
\usepackage{amsmath}
\usepackage{amssymb,bm}
\usepackage{amsthm}
\usepackage{color,dsfont}
\usepackage[colorlinks=true, hyperindex, breaklinks, linkcolor=blue, urlcolor=blue, citecolor=blue]{hyperref} 
\usepackage[normalem]{ulem}
\usepackage[capitalise]{cleveref}
\usepackage{mathrsfs}
\usepackage{mathtools}
\usepackage{amsfonts}
\usepackage{ragged2e}
\usepackage[caption=false]{subfig}
\DeclareCaptionJustification{justified}{\justifying}
\usepackage{lipsum}
\usepackage{float}
\DeclarePairedDelimiter\floor{\lfloor}{\rfloor}

\newcommand{\dual}{{\mathcal{L}^*}}
\newcommand{\dualx}[1]{{\mathcal{L}^*_{#1}}}
\newcommand{\lift}{\texttt{Lift}}
\newcommand{\face}[2]{{\Delta_{#1}(#2)}}
\newcommand{\facex}[3]{{\Delta^{#1}_{#2}(#3)}}
\newcommand{\rest}[1]{|_{#1}}

\renewcommand{\star}[2]{{\mathrm{St}_{#1}(#2)}}

\newcommand{\match}{{\widetilde{\mathcal{G}}}}

\newcommand{\ket}[1]{|#1\rangle}  
\newcommand{\supp}[1]{\ensuremath{\mathrm{supp}\left(#1\right)}}
 
\usepackage{soul}

\theoremstyle{plain}
\newtheorem{theorem}{Theorem}[section]
\newtheorem{lemma}[theorem]{Lemma}

\theoremstyle{definition}

\begin{document}

\title{Triangular color codes on trivalent graphs with flag qubits}

\author{Christopher Chamberland}
\email{christopher.chamberland@ibm.com}
\affiliation{
    IBM T.~J.~Watson Research Center,
    Yorktown Heights, NY, 10598, United States
    }
\author{Aleksander Kubica}
\email{akubica@perimeterinstitute.ca}
\affiliation{
   Perimeter Institute for Theoretical Physics, Waterloo, ON N2L 2Y5, Canada
    }
\affiliation{
   Institute for Quantum Computing, University of Waterloo, Waterloo, ON N2L 3G1, Canada
   }
 \author{Theodore J. Yoder}
\affiliation{
   IBM T.~J.~Watson Research Center,
    Yorktown Heights, NY, 10598, United States
}
\author{Guanyu Zhu}
\affiliation{
   IBM T.~J.~Watson Research Center,
    Yorktown Heights, NY, 10598, United States
    }

\begin{abstract}
The color code is a topological quantum error-correcting code supporting a variety of valuable fault-tolerant logical gates. Its two-dimensional version, the triangular color code, may soon be realized with currently available superconducting hardware despite constrained qubit connectivity. To guide this experimental effort, we study the storage threshold of the triangular color code against circuit-level depolarizing noise. First, we adapt the Restriction Decoder to the setting of the triangular color code and to phenomenological noise. Then, we propose a fault-tolerant implementation of the stabilizer measurement circuits, which incorporates flag qubits. We show how information from flag qubits can be used with the Restriction Decoder to maintain the effective distance of the code. We numerically estimate the threshold of the triangular color code to be $0.2\%$, which is competitive with the thresholds of other topological quantum codes. We also prove that 1-flag stabilizer measurement circuits are sufficient to preserve the full code distance, which may be used to find simpler syndrome extraction circuits of the color code.
\end{abstract}


\maketitle

\section{Introduction}
\label{sec:Intro}

Universal quantum computers offer the exciting potential of solving certain classes of problems with exponential speedups over the best known classical algorithms \cite{Shor94}. However, one of the main drawbacks of quantum devices is their sensitivity to noise. One way to mitigate effects from noise is to fault-tolerantly encode the logical information into error correcting codes, such as stabilizer codes \cite{Gottesman97,CRS96}. Stabilizer codes allow us to extract information about the errors by measuring certain Pauli stabilizer generators without revealing the state of the encoded information. For the special case of topological stabilizer codes \cite{KITAEV97Surface,DKLP02,Bombin2013book}, physical qubits can be placed on a lattice so that the stabilizer generators can be measured using only nearest-neighbor interactions. Information from the stabilizer measurement outcomes (known as the error syndrome) is fed into a classical decoding algorithm whose goal is to find likely error configurations based on the obtained syndrome. We remark that for generic stabilizer codes, the task of optimal decoding is a computationally hard problem \cite{HL11,IyerHarDec15}. Nevertheless, there exist many efficient (but not necessarily optimal) decoding algorithms for topological codes.

In addition to protecting logical information from noise, it is important to perform gates on information encoded in a quantum error correcting code. These gates need to be implemented using fault-tolerant methods to prevent errors from spreading uncontrollably. One simple way to apply logical gates fault-tolerantly is to use transversal operations. 
Unfortunately, there are certain limitations on logical gates implemented by transversal operations \cite{EK09,Zeng2011,Bravyi2013,Pastawski2014,Jochym-OConnor2018}.
In particular, no non-trivial stabilizer code can have a universal logical gate set consisting only of transversal gates.
Still, we should seek quantum codes which not only exhibit good performance in terms of correcting errors but which can also implement as many logical gates transversally as possible. The color code is distinguished in this latter sense since, unlike for the toric code, all logical Clifford gates can be implemented transversally \cite{Bombin06TopQuantDist,Bombin15,KB15, Kubicathesis}.

While computationally advantageous, the color code has lacked competitive decoders and syndrome measurement circuits with two clear problems standing in the way. First, the decoding problem seemed more difficult than decoding the toric code \cite{Wang2010,Landahl2011}, which can be solved via a simple matching algorithm.
Second, color code stabilizer generators are higher weight than the toric code stabilizers, thus in general requiring more ancilla qubits.
The first problem has been extensively studied \cite{Bombin2012a,Sarvepalli2012,Aloshious2016,Brown2015,ProjectionDecoder,LinTimeDec17}.
The thresholds for optimal correction, obtained by statistical-mechanical mappings, are very comparable for both the toric and color codes \cite{DKLP02,Katzgraber2009,Bombin2012,KubicaStatMech}. Thus, one should not expect the inferior error-correction performance of the color code. Indeed, this was confirmed with color code decoders matching the performance of the toric code decoders \cite{MKJO19, Tuckett2018,KD19} assuming perfect syndrome measurement circuits.

Dealing with the second problem, thus extending the competitiveness of the color code to circuit-level noise, is the subject of the current paper. Our starting point is the recently proposed decoder for the color code, the Restriction Decoder~\cite{KD19}, which is particularly appealing due to its simplicity and good performance. The Restriction Decoder builds on the close connection between the toric and color codes in any dimension~\cite{Kubica2015}. It combines any toric code decoder with a local lifting procedure to find a color code correction.
Importantly, the Restriction Decoder seems to be a good candidate for adaptation to realistic circuit-level noise.

\begin{table*}[ht!]
	\begin{centering}
		\begin{tabular}{|c|c|c|c|c|}
			\hline
Topological  code& Connectivity &  Total number of qubits & Transversal Gates &  Threshold \\ \hline\hline
Rotated  surface code & $\{ 4,4 \}$  & $2d^2 - 1$ & $\overline{X}$, $\overline{Z}$, $\overline{\text{CNOT}}$ &  $p_{th} \approx 0.7\%$ \cite{TC_threshold} \\ \hline
\begin{tabular}{@{}c@{}} *Heavy Hexagon code  \\ \emph{(only $X$ errors)} \end{tabular}
& $\{12/5 ,3\}$   & $(5d^2 - 2d - 1)/2$ & $\overline{X}$, $\overline{Z}$, $\overline{\text{CNOT}}$ & $p_{th} \approx 0.45\%$  \cite{CZYHC19} \\ \hline
Heavy Square code  & $\{8/3, 4\}$ & $3d^2 - 2d$ & $\overline{X}$, $\overline{Z}$, $\overline{\text{CNOT}}$ & $p_{th} \approx 0.3\%$  \cite{CZYHC19} \\ \hline
Triangular color code & $\{ 3, 3\}$  & $(3d-1)^2/4$ & $\overline{S}, \overline{H}, \overline{\text{CNOT}}$& $p_{th} \approx 0.2\%$  \\ \hline
		\end{tabular}
		\par\end{centering}		
	\caption{\label{Tab:ThreshCompare}
A comparison of the triangular color code with other topological codes: the rotated surface code \cite{BK98,TS14} and the Heavy Hexagon/Square codes \cite{CZYHC19}.
The connectivity $\{c_1 , c_2 \}$ denotes the average number $c_1$ of nearest neighbor qubits (qubits connected by a CNOT gate) and the maximum number $c_2$ of qubits interacting with any qubit in a fault-tolerant implementation of the code.
We point out that a smaller average connectivity results in fewer frequency collisions and cross-talk errors for superconducting qubit architectures \cite{CZYHC19}.
We provide the total number of qubits (including the data, syndrome measurement and flag qubits) as a function of the code distance $d$, as well as the generating set of transversal gates, where $X$, $Z$, $S = \text{diag}(1,i)$, $H$ and CNOT are the Pauli $X$ and $Z$, phase, Hadamard and controlled-NOT gates (a bar above the gate denotes the logical version of the gate).
Note that $S$, $H$ and CNOT generate the Clifford group.
We observe that the storage thresholds of the analyzed codes against the circuit-level noise model (described in \cref{subsubEdgeRen}) are comparable.
\emph{*The threshold of $0.45\%$ for the Heavy Hexagon code is only for one type of Pauli error, since the code is a hybrid surface/Bacon-Shor code, and the Bacon-Shor code has no threshold \cite{AliferisCross07}.}
{}
	}
	\end{table*}

In this work, we obtain three key results.
Our first result focuses on the adaptation of the Restriction Decoder to the triangular color code and to phenomenological noise. We emphasize that the original version of the Restriction Decoder is applicable only to the color code on a lattice without boundaries and to the scenario of perfect syndrome extraction, i.e., when there are no measurement errors. A naive adaptation of the Restriction Decoder would be to perform the decoding of the toric code with boundaries, and then apply a generalized lifting procedure also to the boundary. This naive adaptation would, however, result with the effective distance dropping by half (i.e. the code would correct errors arising from $\floor{(d-3)/4}$ faults). In our adaptation of the Restriction Decoder, we thus address this problem by first finding connected components of the error syndrome, and only then applying the local lifting procedure. 
We numerically verify that for codes with distance $d=5,7,9$ such a scheme corrects any error of weight at most $2$, $2$ and $3$, respectively.
Moreover, for any code with distance $d = 6n+1$, where $n$ is a positive integer, we find an error of weight $2n+1$, which leads to a logical error; see Appendix~\ref{sec_naive}.
Thus, we expect the adapted Restriction Decoder to correct any error up to weight $\sim d/3$.

As was shown in Ref.~\cite{CZYHC19}, for superconducting qubit architectures with fixed frequency qubits, frequency collisions and cross-talk errors can be reduced by using codes where the degree of the connectivity between qubits is small. For our second result, we show how the color code qubits and \emph{all} of the required ancilla qubits can be placed on vertices of the hexagonal lattice such that syndrome measurement circuits interact only with neighboring qubits. Consequently, each qubit has degree three connectivity. Further, we show how the syndrome measurement circuits take advantage of additional ancilla qubits used as flag qubits \cite{CR17v1,CR17v2,CB17,TCD18Flag,ReichardtFlag18,ChamberlandMagic,ChamberlandGKP,CZYHC19}. In an error correction scheme, the role of flag qubits\footnote{We point out that in \cite{DA07}, unverified cat states interacting with data qubits, along with a decoding circuit can be used to fault-tolerantly measure stabilizers. However such methods add additional overhead compared to the flag schemes presented in this paper. Further, it is not known if they can be implemented in a decoding scheme in a scalable and efficient way.} is to detect errors of weight greater than $v$ arising from $v < d_{\text{eff}}$ faults (where $d_{\text{eff}}$ if the effective distance of the code with a given decoder\footnote{Note that given an error correcting code of distance $d$ and a decoder used to correct errors, it is possible that not all errors arising from $\floor{(d-1)/2}$ faults can be corrected (in many cases a suboptimal decoder is preferred for reasons of speed and scalability). In such a case $d_{\text{eff}} < d$. }) and to provide additional information allowing such errors to be identified and corrected. Further, we show how to supplement the Restriction Decoder with information from the flag qubits to maintain the full effective distance of the color code. In particular, we show that using 1-flag circuits is enough to recover the effective distance of the code. We emphasize that in our proof the structure of the color code plays an important role, since weight-three Pauli errors of the same type arising from two faults in a stabilizer measurement circuit cannot have full support along a minimum-weight logical operator of the color code.

For our third result, we provide numerical estimates of the storage threshold of the adapted Restriction Decoder for the triangular color code against both code capacity and a full circuit-level depolarizing noise model.
We estimate the Restriction Decoder threshold for the triangular color code against the circuit-level depolarizing noise model to be 0.2\%. In \cref{Tab:ThreshCompare}, we compare the connectivity, total number of qubits, transversal gate sets and threshold of the triangular color code with other topological codes.
We remark that the triangular color code threshold for the circuit-level depolarizing noise model without  the use of flag qubits and edge weight renormalization has also been independently investigated in Refs.~\cite{Beverlandthesis,BKS19}. We also point out that in \cite{BCCBO9}, the color code with flag qubits was studied for small distances using machine learning methods where no prior information regarding the noise model is required. Although such methods lead to good code performance compared to algorithmic decoders, they can only be applied to small distance codes since they are not scalable \cite{CRMachineLearning18}.

The manuscript is organized as follows. \cref{sec:DecoSec} is devoted to decoding the color code. We begin in \cref{subsec:TriangColorRev} by briefly reviewing the triangular color code and in \cref{subsec:DecColorRev}, we review the Restriction Decoder for the color code. We then show how to adapt the Restriction Decoder to the triangular color code in \cref{subsec:AdaptDecTriangColor}, and in \cref{Subsec:IncMeasErrs} we show how to incorporate measurement errors into the decoder. In \cref{sec:FTColorCodeCircuits}, we focus on the fault-tolerant implementation of the color code on low degree graphs. In \cref{subsec:CircuitLayout}, we describe the circuit layout of the triangular color code on a graph of degree three. We then provide the syndrome measurement circuits in addition to the CNOT gate scheduling for a full round of stabilizer measurements. In \cref{subsec:FlagQubitInfo}, we show how flag qubits can be used to correct high weight errors arising from fewer faults, and prove in \cref{app:more_thm} that only 1-flag circuits are required to measure the weight-six stabilizers. In \cref{subsub:FlagEdges} we show how to add flag edges to the decoding graphs, and in \cref{subsubEdgeRen} we show how edges of the graph are renormalized based on the flag qubit measurement outcomes.  We conclude \cref{sec:FTColorCodeCircuits} by providing an alternative flag scheme which does not require modifications to the decoding graphs in \cref{subsubsec:AltFlag}. In \cref{sec:NumRes}, we provide numerical results of the performance of the triangular color code under both code capacity and circuit level depolarizing noise models. We conclude in \cref{sec:Conclusion} and discuss future work.

\section{Color code decoding}
\label{sec:DecoSec}

\subsection{Triangular color code}
\label{subsec:TriangColorRev}

The triangular color code is a version of the color code defined on a two-dimensional lattice $\mathcal{L}$ with three boundaries, see \cref{fig_stuff}(a).
We choose the lattice $\mathcal{L}$ to be a finite region of the hexagonal lattice.
Importantly, the lattice $\mathcal{L}$, as a color code lattice, has to satisfy the following two conditions 
\begin{itemize}
\item 3-valence: all the vertices except for the three corner vertices of $\mathcal{L}$ are incident to three edges,
\item 3-colorability: every face of $\mathcal{L}$ can be colored in one of three colors $R$, $G$ and $B$ in such a way that any two neighboring faces sharing an edge have different colors.
\end{itemize}
We place one qubit at each vertex of $\mathcal{L}$.
For each face of $\mathcal{L}$ we introduce both $X$- and $Z$-type stabilizer generators, each of which are supported on all the qubits belonging to that face.
The code space is defined as the $(+1)$-eigenspace of all the stabilizer generators.
We remark that any logical Pauli operator of the color code can be implemented as a tensor product of Pauli operators supported within a 1D string-like region.

The discussion of the color code decoding can be simplified if we use a lattice $\dual$ dual to the lattice $\mathcal{L}$.
We illustrate the dual lattice $\dual$ in \cref{fig_stuff}(b).
Since $\mathcal{L}$ satisfies the 3-valence condition, $\mathcal{L}^*$ necessarily consists of triangular faces.
By definition, two faces of $\dual$ share an edge if and only if the corresponding two vertices of $\mathcal{L}$ are incident to the same edge of $\mathcal{L}$.
Lastly, the vertices of $\dual$ correspond to faces and the three boundaries of $\mathcal{L}$.
Thus, the vertices of $\dual$ endow the colors of the corresponding faces (or boundaries) of $\mathcal{L}$.
Note that in the dual lattice $\dual$, qubits are placed on faces and stabilizer generators are identified with vertices (except for the three boundary vertices $v_R$, $v_G$ and $v_B$).

\subsection{Color code decoding problem}
\label{subsec:DecColorRev}

Since color code stabilizer generators can be chosen to be either $X$- or $Z$-type, we choose to independently correct the bit-flip $X$ and phase-flip $Z$ errors.
Moreover, the 2D color code is self-dual, i.e., $X$- and $Z$-type stabilizer generators have the same support, implying that the procedures of correcting $X$ and $Z$ errors are analogous.
Thus, in what follows we focus our discussion only on $X$ errors.

\begin{figure*}
\centering
(a)\includegraphics[width=0.37\textwidth]{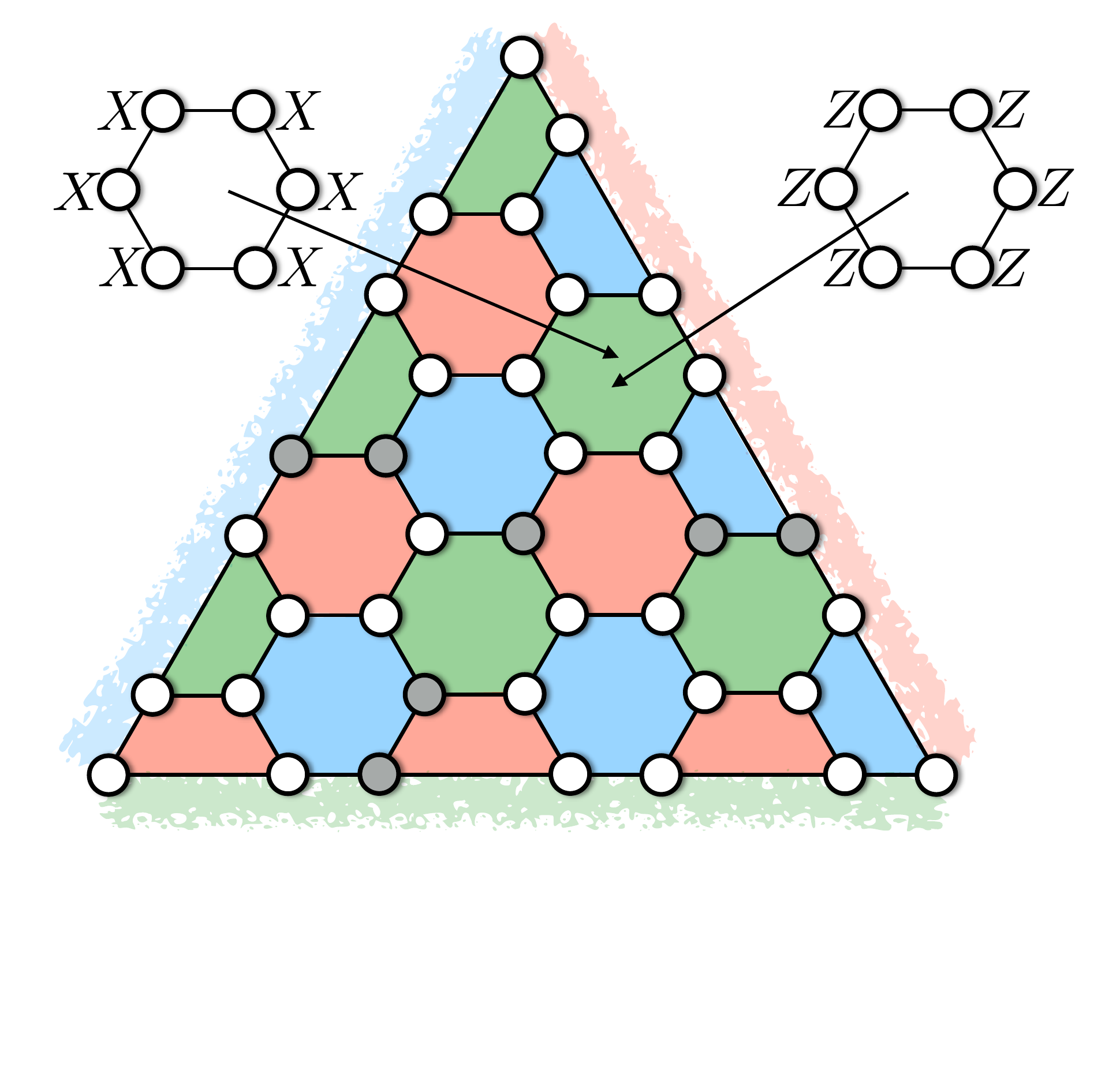}\hspace*{15mm}
(b)\includegraphics[width=0.37\textwidth]{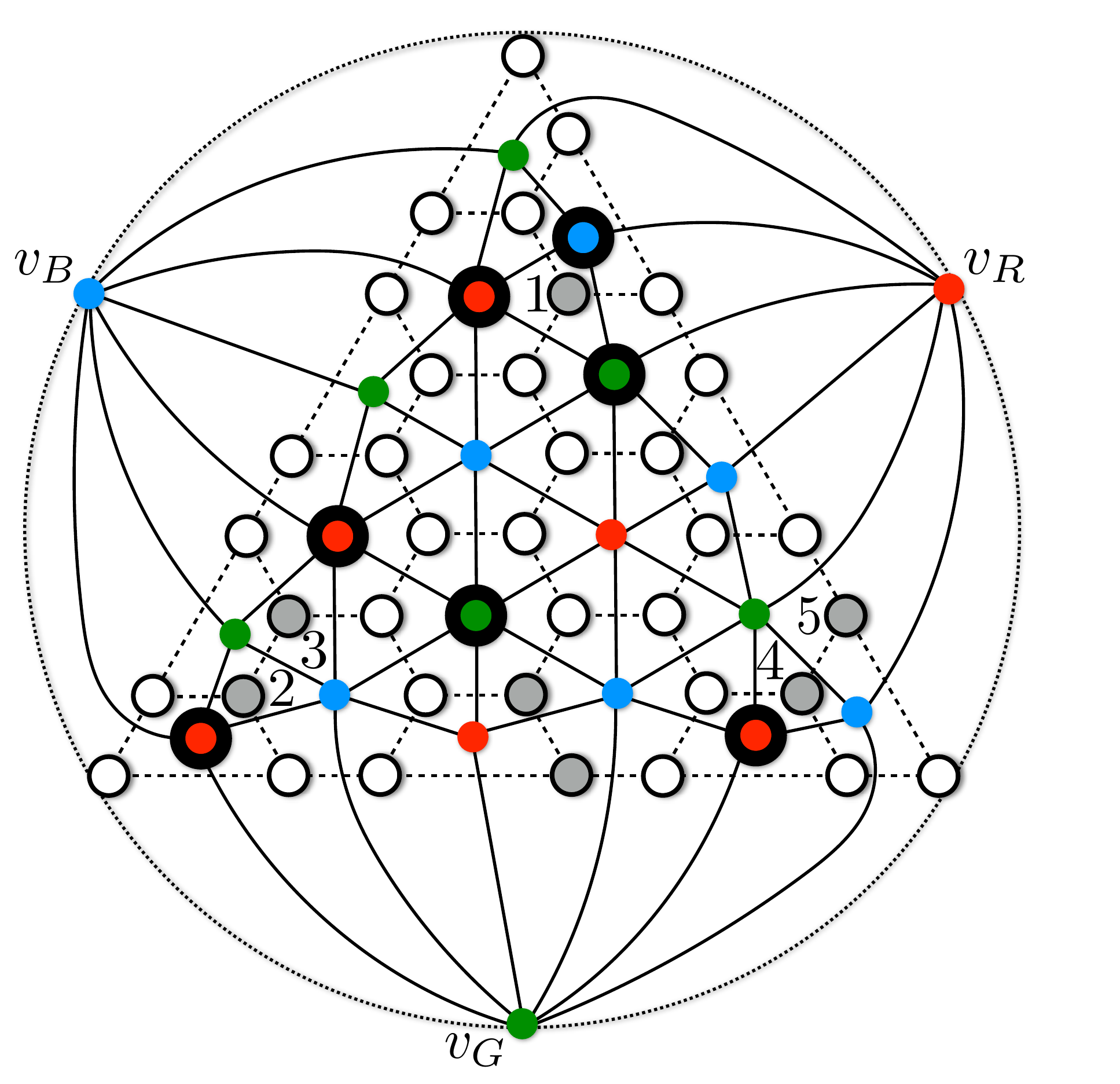}\\
(c)\includegraphics[width=0.37\textwidth]{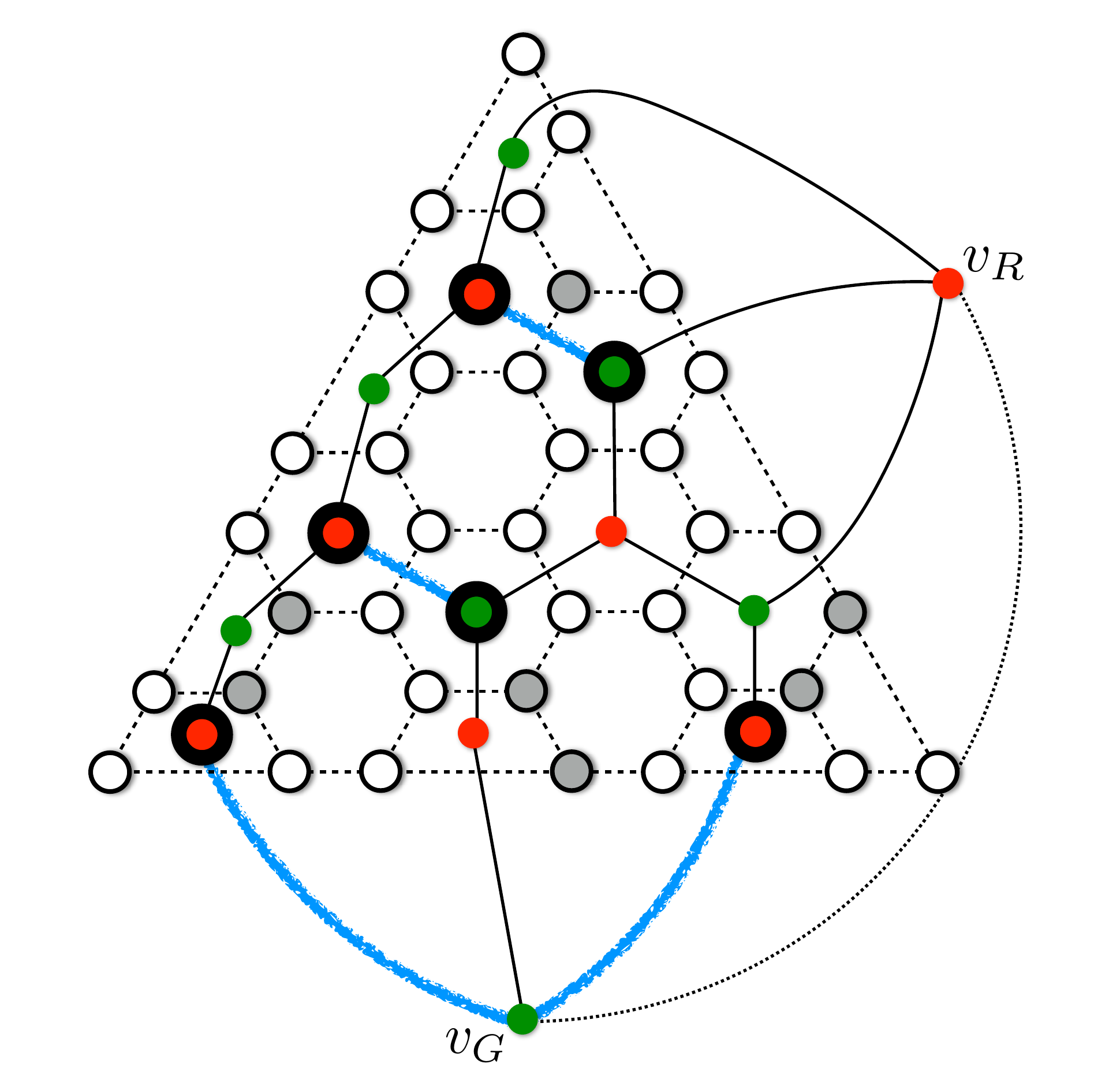}\hspace*{15mm}
(d)\includegraphics[width=0.37\textwidth]{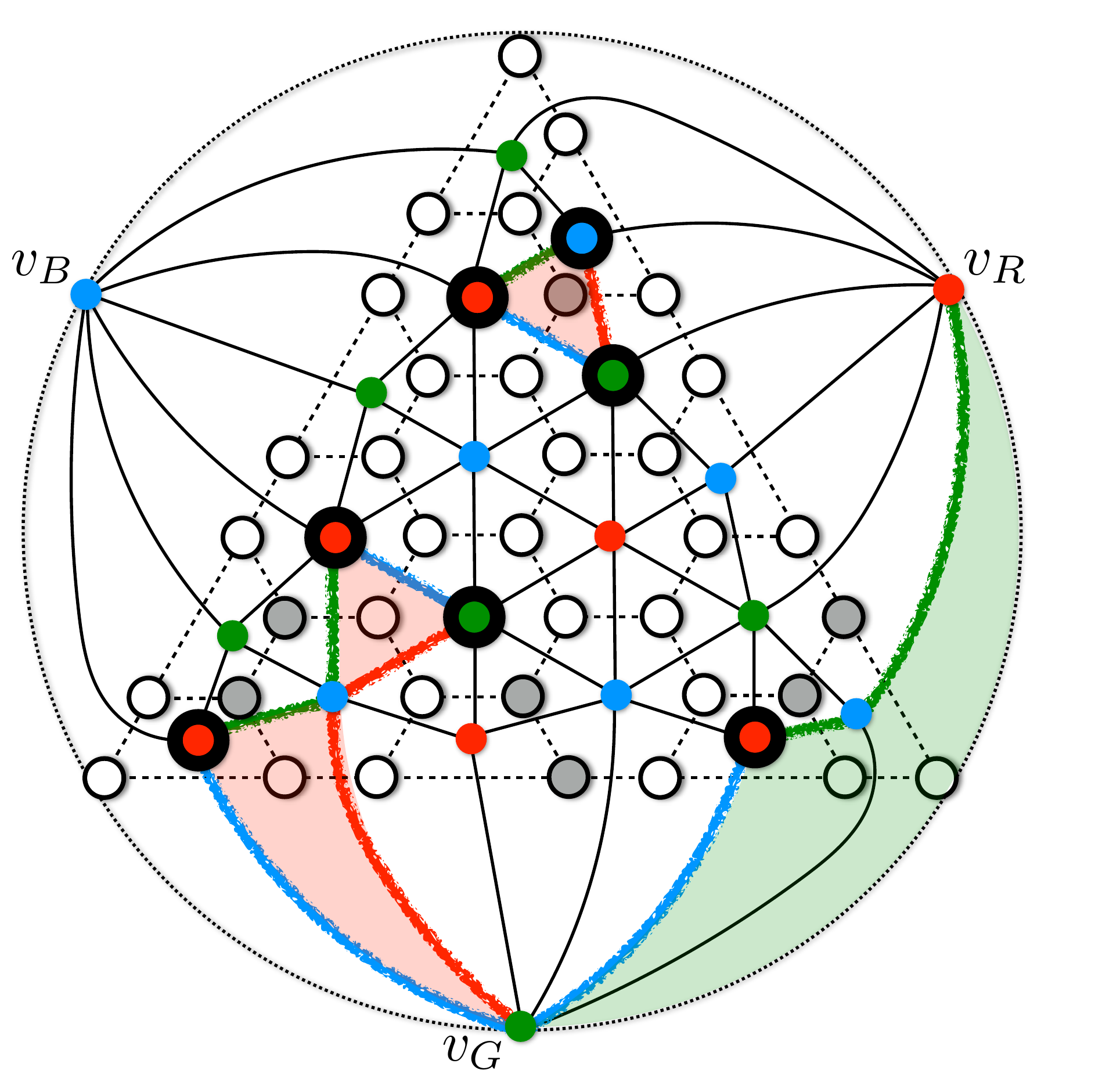}
\caption{
(a) The triangular color code defined on a two-dimensional lattice $\mathcal{L}$, which is a region of the hexagonal lattice.
Qubits are placed on vertices of $\mathcal{L}$, whereas $X$- and $Z$-type stabilizers are associated with faces of $\mathcal{L}$.
Logical Pauli $X$ and $Z$ operators can be supported within a 1D string-like region (shaded in gray).
(b) The dual lattice $\dual$ with the boundary vertices $v_R$, $v_G$ and $v_B$.
A subset of qubits $\epsilon$ affected by Pauli $X$ errors (shaded in gray), which anticommute with some stabilizer generators $\sigma$ (marked red, green and blue vertices).
(c) The restricted lattice $\dualx{RG}$ is obtained from $\dual$ by removing all the $B$ vertices, as well as all the edges and faces incident to them.
We find the subset of edges $\rho_{RG}\subseteq\face 1 {\dualx{RG}}$ (blue lines) by pairing up the vertices of the restricted syndrome $\sigma_{RG}$. 
(d) The adapted Restriction Decoder finds a correction $\phi(\sigma)$ (qubits on shaded triangular faces) by first combining the parings $\rho_{RG}$, $\rho_{RB}$ and $\rho_{GB}$, then finding connected components of the syndrome, and finally applying the local lifting procedure \lift.
Note that the error $\epsilon$ and the correction $\phi(\sigma)$ differ by some stabilizer operator.
}
\label{fig_stuff}
\end{figure*}

Let us denote by $\face 0 \dual$, $\face 1 \dual$ and $\face 2 \dual$ the sets of vertices, edges and faces of the dual lattice $\dual$.
For convenience, we exclude the edges $(v_R,v_G)$, $(v_R,v_B)$ and $(v_G,v_B)$ from $\face 1 \dual$.
Let $\epsilon\subseteq \face 2 {\dual}$ be the subset of faces, which correspond to qubits affected by $X$ errors.
The corresponding syndrome $\sigma\subseteq \face 0 \dual \setminus \{ v_R, v_G, v_B\}$ is the subset of all the $Z$-type stabilizer generators, which anticommute with the error $\epsilon$.
The syndrome $\sigma$ can be found as the subset of vertices, which are incident to an odd number of faces in $\epsilon$.
As illustrated in \cref{fig_stuff}(b), a single $X_1$ error anticommutes with three neighboring stabilizer generators, whereas a two-qubit $X_2X_3$ error anticommutes with two stabilizer generators.
Note that near the boundary a two-qubit error $X_4 X_5$ anticommutes with only one stabilizer generator.

The problem of color code decoding can now be formulated as follows:
given the $Z$-type syndrome $\sigma\subseteq\face 0 \dual \setminus \{ v_R, v_G, v_B\}$, find an $X$-type correction operator supported on $\phi\subseteq\face 2 \dual$, whose syndrome matches $\sigma$.
We can view color code decoding as a task of finding a subset of faces $\phi$ from some subset of vertices $\sigma$.
Importantly, $\phi$ has to satisfy the following condition: 
a vertex $v\in\face 0 \dual \setminus \{ v_R, v_G, v_B\}$ is incident to an odd number of faces of $\phi$ iff $v$ belongs to $\sigma$.
We say that decoding of the error $\epsilon$ succeeds iff $\epsilon$ combined with the correction $\phi$ forms a trivial logical operator.
In the rest of this section, we describe a novel decoder for the triangular color code which builds upon the recently introduced Restriction Decoder~\cite{KD19}.

\subsection{Adaptation of the Restriction Decoder}
\label{subsec:AdaptDecTriangColor}

First, following Ref.~\cite{KD19}, we review a couple of concepts, such as boundaries and restricted lattices.
Let $\alpha\subseteq \face 2 \dual$ and $\beta\subseteq \face 1 \dual$ be some subsets of faces and edges of $\dual$.
We denote by $\partial_2\alpha\subseteq \face 1 \dual$ and $\partial_1\beta\subseteq \face 0 \dual$ the sets of all the edges and vertices of $\dual$, which belong to an odd number of faces in $\alpha$ and edges in $\beta$, respectively.
We refer to $\partial_2\alpha$ and $\partial_1\beta$ as the $1$-boundary of $\alpha$ and $0$-boundary of $\beta$.
We construct the restricted lattice $\dualx{RG}$ by removing from $\dual$ all the vertices of color $B$ as well as all the edges and faces incident to the removed vertices; see \cref{fig_stuff}(c).
In other words, $\dualx{RG}$ contains $R$ and $G$ vertices, as well as edges between them; the restricted lattices $\dualx{RB}$ and $\dualx{GB}$ are defined analogously.
Lastly, we denote by $\sigma_{RG}\subseteq\face 0 {\dualx{RG}}\setminus \{ v_R, v_G\}$ the set of all the $R$ and $G$ vertices of the syndrome $\sigma\subseteq\face 0 \dual\setminus \{ v_R, v_G, v_B\}$; we define $\sigma_{RB}$ and $\sigma_{GB}$ in a similar way.

The first step of the adaptation of the Restriction Decoder to the triangular color code is to pair up vertices of $\sigma_C$ within the restricted lattice $\dualx C$, where $C \in \{ RG, RB, GB\}$.
This step, roughly speaking, allows us to find a subset of edges $\rho_C\subseteq\face 1 {\dualx C}$ with the $0$-boundary matching $\sigma_C$.
For instance, for every vertex in $\sigma_{RG}$ we either choose to pair it up with another vertex in $\sigma_{RG}$ or with the boundary vertex $v_R$; see \cref{fig_stuff}(c).
Then, we can find a subset of edges $\rho_{RG}\subseteq\face 1 {\dualx{RG}}$, such that the $0$-boundary of $\rho_{RG}$ is 
\begin{equation}
\partial_1 \rho_{RG} = \sigma_{RG} \cup U,
\end{equation}
where $U$ is some (possibly empty) subset of $\{ v_R, v_G \}$.
Similarly, we find $\rho_{RB}\subseteq\face 1 {\dualx{RB}}$ and $\rho_{GB}\subseteq\face 1 {\dualx{GB}}$, whose $0$-boundaries can only differ from $\sigma_{RB}$ and $\sigma_{GB}$ by some subsets of $\{ v_R, v_B \}$ and $\{ v_G, v_B \}$, respectively.

Before proceeding, we need to introduce the notion of a connected component (see \cref{fig_stuff}(d)).
We say that two different vertices $u,w\in\sigma\cup\{v_R,v_G,v_B\}$ are connected iff they have been paired up within at least one of the three restricted lattices.
We say that a subset of syndrome vertices $s=\{v_1,\ldots,v_{n}\}\subseteq\sigma$ forms a connected component iff there exist two (possibly the same) boundary vertices $v_0,v_{n+1}\in \{v_R,v_G,v_B\}$, such that $v_{i-1}$ and $v_i$ are connected for any $i=1,\ldots,n+1$, where $n\geq 1$.
We define the pairing
\begin{equation}
\gamma(s)\subseteq \rho_{RG}\cup\rho_{RB}\cup\rho_{GB}
\end{equation}
of the connected component $s$ to be the subset of edges of $\rho_{RG}\cup\rho_{RB}\cup\rho_{GB}$ used to pair up $v_{i-1}$ with $v_i$.

In the second step of the adapted Restriction Decoder we find the set $\Sigma$ of all the connected components $s_1,\ldots,s_j \subseteq \sigma$, which are connected to the boundary vertex $v_R$.
For each connected component $s\in\Sigma$ let us denote by $v_0$ and $v_{n+1}$ the two boundary vertices the connected component $s$ is connected to.
Note that, by definition, one of the boundary vertices $v_0$ and $v_{n+1}$ has to be $v_R$.
If both $v_0$ and $v_{n+1}$ have color $R$, then we set\footnote{
This choice is arbitrary and we can set $C(s) = B$.}
the color $C(s) = G$.
Otherwise, we set $C(s)$ to be one of three colors $R$, $G$ and $B$, which is different from the colors of $v_0$ and $v_{n+1}$.

Now, we recall a couple of notions introduced in Ref.~\cite{KD19}.
Let $\beta\subseteq\face 1 \dual$ be some subset of edges of $\dual$.
We define $\facex C 0 \beta$ to be the set of all the vertices of color $C\in\{R, G, B\}$, which belong to at least one edge of $\beta$, and denote by $\beta\rest v$ the set of all the edges of $\beta$ incident to a vertex $v\in\face 0 \dual$.
We also define $\star 2 v$ to be the set of all the faces of $\dual$ incident to the vertex $v$.

The third step of the adapted Restriction Decoder applies a local lifting procedure \lift\ to some of the vertices of $\dual$.
Roughly speaking, the local lifting procedure \lift\ allows us to find a subset of faces incident to the same vertex, whose $1$-boundary locally matches either
\begin{equation}
\rho = (\rho_{RG}\cup\rho_{RB})\setminus\bigcup_{s\in\Sigma} \gamma(s)
\end{equation}
or the pairing $\gamma(s)$ of one of the connected components $s\in\Sigma$.
To be more precise, for every vertex $v\in\facex R 0 {\rho}$ or $v\in V(\Sigma)$, where
\begin{equation}
V(\Sigma) = \bigcup_{s\in\Sigma}\facex {C(s)} 0 {\gamma(s)},
\end{equation}
we apply \lift\ to find a subset of faces $\tau_v \subseteq\star 2 v$, such that the $1$-boundary of $\tau_v$ matches either $\rho$ or $\gamma(s)$ in the neighborhood of $v$, i.e.,
$(\partial_2 \tau_v)\rest v = \rho\rest v$ or $(\partial_2 \tau_v)\rest v = \gamma(s)\rest v$, respectively.

Finally, the adapted Restriction Decoder returns
$\phi(\sigma) = \bigcup_{v\in\facex R 0 {\rho} \cup V(\Sigma)} \tau_v$
as a correction for the syndrome $\sigma$.
One can show that the syndrome of $\phi(\sigma)$ is $\sigma$, thus $\phi(\sigma)$ is indeed a valid correction of the error $\epsilon$.
We illustrate the correction found by the adapted Restriction Decoder in \cref{fig_stuff}(d).

To summarize, the adapted Restriction Decoder consists of the following steps.
\begin{enumerate}
\item For every color $C\in\{ RG, RB, GB\}$ pair up the restricted syndrome $\sigma_C$ within the restricted lattice $\dualx C$ to find $\rho_C$, whose $1$-boundary matches $\sigma_C$.
\item Find the set $\Sigma$ of all the connected components $s_1,\ldots,s_j$ of the syndrome $\sigma$, which are connected to the boundary vertex $v_R$.
\item Apply the local lifting procedure \lift\ to every vertex $v \in {\facex R 0 {\rho} \cup V(\Sigma)}$ to find a subset of faces $\tau_v$, whose $1$-boundary locally matches $\rho$ or $\gamma(s_i)$.
\item Return the correction $\phi(\sigma) = \bigcup_{v\in\facex R 0 {\rho} \cup V(\Sigma)} \tau_v$.
\end{enumerate}

We would like to make a couple of remarks about the adapted Restriction Decoder.
\begin{itemize}
\item[(i)] The local lifting procedure \lift\ can naively be implemented in constant time by checking for all possible subsets $\tau_v\subseteq\star 2 v$ of faces incident to the given vertex $v$ whether the $1$-boundary of $\tau_v$ locally matches $\rho$ or $\gamma(s)$.
\item[(ii)] The complexity of the adapted Restriction Decoder is determined by the complexity of finding the pairing of the restricted syndrome $\sigma_C$ within the restricted lattice $\dualx{C}$ for $C\in \{RG, RB, GB \}$.
This problem, in turn, can be efficiently solved using e.g. the Minimum Weight Perfect Matching (MWPM) algorithm \cite{Edmonds65}.
\item[(iii)] Step 2 of the adapted Restriction Decoder is the main difference from the original version of the Restriction Decoder in Ref.~\cite{KD19}.
Namely, in the presence of the boundaries we find connected components of the syndrome and then apply the local lifting procedure \lift\ to certain vertices along the pairing for each connected component.
\item[(iv)] 
We expect the adapted Restriction Decoder to be able to correct all errors of weight at most $\sim d/3$.
We numerically found that in the sub-threshold regime
the scaling of the logical error rate is well described by $p_{L} \sim ap^{d/3}$.
However, in practice the leading order coefficient is several orders of magnitude smaller than the coefficient of the next order term.
A plot of the logical $Z$ error rate is given in \cref{fig:2DLogZPlot} (see \cref{sec:NumRes}).
We illustrate an example of a smallest weight error leading to a logical error in \cref{fig_smallest_errors}(b) (see \cref{sec_naive}).
\item[(v)] A naive generalization of the Restriction Decoder could treat the boundary vertex $v_R$ on the same footing as any $R$ vertex in the bulk of the lattice $\dual$.
In particular, one could first pair up the restricted syndromes $\sigma_{RG}$ and $\sigma_{RB}$ within the restricted lattices $\dualx{RG}$ and $\dualx{RB}$, and then apply the local lifting procedure \lift\ to every $R$ vertex of $\dual$, including the boundary vertex $v_R$.
This decoder, however, is only guaranteed to correct errors of weight at most $\lfloor\frac{d-3}{4}\rfloor$ for odd $d$.
We discuss such a naive generalization of the Restriction Decoder in~\cref{sec_naive}.
\end{itemize}

\subsection{Incorporating measurement errors}
\label{Subsec:IncMeasErrs}

Until now, we have assumed that the syndrome $\sigma$ can be extracted perfectly, i.e., the stabilizer measurement circuits do not introduce any errors into the data qubits and there are no measurement errors.
In the remainder of this section, we explain how one can use the adapted Restriction Decoder in the presence of measurement errors.
In such a setting, to get a reliable estimate of the syndrome we need to repeat stabilizer measurements $T$ times, where $T$ is comparable with the code distance.
Let us denote by $\sigma_t \subseteq \face 0 {\dual}$ the (possibly faulty) syndrome extracted at time step $t = 0,1,\ldots,T$.
We use the syndrome information collected over time to find an appropriate correction.

First, let us introduce the matching graph $\mathcal{G}_C = (V_{\mathcal{G}_C}, E_{\mathcal{G}_C})$ for any pair of colors $C\in\{RG,RB,GB\}$.
The vertices of the matching graph $\mathcal{G}_C$ correspond to the vertices of the restricted lattice $\dualx C$, i.e., $V_{\mathcal{G}_C} = \face 0 {\dualx C}$.
The set of edges of $\mathcal{G}_C$ contains not only all the edges of $\dualx C$, i.e., $E_\mathcal{G}\supset\face 1 {\dualx C}$, but also certain flag edges; see~\cref{fig:AllRenormFlags}.
A flag edge is added for any two vertices of $V_{\mathcal{G}_C}$ of the same color, whose graph distance in the restricted lattice $\dualx C$ is two.
To give the reader some intuition, flag edges are added to the matching graph to capture the possibility of a weight-two data qubit error being introduced by a single fault in a stabilizer measurement circuit\footnote{
In \cref{subsec:FlagQubitInfo}, we show that flag edges for weight-three data qubit errors arising from two faults in a stabilizer measurement circuit are not required.}.
We defer the detailed discussion of flag edges to \cref{subsub:FlagEdges}.

Now, we can construct the space-time matching graph $\match_{C} = (V_{\match_C},E_{\match_C})$ for any pair of colors $C\in\{RG,RB,GB\}$.
The vertices of the space-time matching graph $\match_C$ correspond to the elements of the set $V_{\mathcal{G}_C} \times \{ 1,\ldots,T\}$.
Two vertices $(u,t)$ and $(w,t)$ of the space-time matching graph $\match_C$ are connected by an edge in $E_{\match_C}$, where $t=1,\ldots, T$ and $u,w\in V_{\mathcal{G}_C}$, whenever two vertices $u$ and $w$ of the matching graph $\mathcal{G}_C$ are connected by an edge, i.e., $(u,w) \in E_{\mathcal{G}_C}$.
Moreover, for $t=1,\ldots,T-1$ two vertices $(u,t),(w,t+1)\in V_{\match_C}$ are connected by an edge $e\in E_{\match_C}$ iff either $u=w$ or $e$ corresponds to a diagonal edge.
Diagonal edges are added to the space-time matching graph $\match_C$ to account for space-time correlated errors introduced into the data qubits by two-qubit gate failures occurring in the stabilizer measurement circuits.
A detailed description of diagonal edges is provided in \cref{app:EdgeWeightCalc}.
We remark that the edges of the space-time matching graph are assigned weights, which depend on the stabilizer measurement circuits and the flag measurement outcomes; see \cref{sec:FTColorCodeCircuits,app:EdgeWeightCalc} for more information.

Let us denote by $\mathbb{F}_2(A)$ a vector space over $\mathbb{F}_2$, whose basis corresponds to the elements of some finite set $A$.
Since there is a one-to-one correspondence between the vectors in $\mathbb{F}_2(A)$ and the subsets of $A$, we would treat them interchangeably.
For $C\in\{RG,RB,GB\}$ let $e\in E_{\match_C}$ be an edge of the space-time matching graph $\match_C$ connecting two vertices
$(v_1,t_1),(v_2,t_2) \in V_{\match_C}$.
We define $f:\mathbb{F}_2(E_{\match_C})\rightarrow \mathbb{F}_2(E_{\mathcal{G}_C})$ as a linear map defined on every basis element $e$ by
\begin{equation}
f(e) = \begin{cases}
(v_1,v_2)&\quad\textrm{if $t_1 = t_2$ or $e$ is a diagonal edge},\\
0&\quad\textrm{otherwise,}
\end{cases}
\end{equation}
where $(v_1,v_2)$ denotes the edge connecting two vertices $v_1$ and $v_2$ in the matching graph $\mathcal{G}_C$.
We refer to $f$ as the flattening map.
Moreover, we define $g:\mathbb{F}_2(E_{\mathcal{G}_C})\rightarrow \mathbb{F}_2(\face 1 \dual)$ as a linear map defined on every basis element $e \in E_{\mathcal{G}_C}$ by
\begin{equation}
g(e) = \begin{cases}
e&\quad\textrm{if $e\in\face 1 {\dualx C}$},\\
e_1\cup e_2 &\quad\textrm{if $e$ is a flag edge},\\
\end{cases}
\label{eq:Flattg}
\end{equation}
where (whenever $e$ is the flag edge) $e_1$ and $e_2$ are the two edges in the restricted lattice $\dualx C$ connecting the endpoints of $e$ (see \cref{sec:FTColorCodeCircuits} and in particular \cref{fig:FlagEdges}).
We refer to $g$ as the flag projection map.  

Lastly, we define the set of highlighted vertices $W\subset \face 0 {\dual} \times \{1,\ldots,T\}$ as follows
\begin{equation}
W = \bigcup_{t=1}^T (\sigma_{t-1} \oplus \sigma_{t})\times \{ t \},
\end{equation}
where $\oplus$ denotes the symmetric difference of the syndromes $\sigma_{t-1}$ and $\sigma_{t}$ extracted at time steps $t-1$ and $t$.
We can think of the vertices $W$ as the space-time locations, where stabilizer measurement outcomes change in between two consecutive measurement rounds.
The restricted highlighted vertices $W_{RG}$ are defined as the subset of the highlighted vertices $W$ within the space-time matching graph $\match_{RG}$; similarly $W_{RB}$ and $W_{GB}$.
Also, we would refer to the set $\partial V = \{v_R, v_G, v_B \} \times \{1,\ldots, T\} \subset V_\match$ as the space-time boundary vertices of the space-time matching graph $\match$.

Now we are ready to describe how the Restriction Decoder can be used for the triangular color code in the presence of measurement errors.
First, we pair up the restricted highlighted vertices $W_C$ within the space-time matching graph $\match_C$, where $C\in\{ RG, RB, GB\}$, 
and find a subset $\widetilde\rho_C$ of the edges of $\match_C$, whose $0$-boundary matches (up to the space-time boundary vertices $\partial V$) $W_C$.
Then, we find the set $\widetilde\Sigma$ of all the connected components $\tilde s_1,\ldots,\tilde s_j \subseteq W$, which are connected to any of the space-time boundary vertices $(v_R,t)\in\partial V$, where $t=1,\ldots,T$.
Next, we apply the flattening map $f$ followed by the flag projection map $g$ to the pairing $\tilde\gamma(\tilde s)$ of each connected component $\tilde s\in \widetilde\Sigma$ and to
$\widetilde\rho = (\tilde\rho_{RG}\cup \tilde\rho_{RB})\setminus\bigcup_{\tilde s\in\tilde\Sigma} \tilde\gamma(\tilde s)$.
Then, we apply the local lifting procedure \lift\ to certain vertices of $g\circ f (\tilde s)$ and $g\circ f (\widetilde\rho)$ for $\tilde s\in \widetilde\Sigma$ in an analogous way as described in \cref{subsec:AdaptDecTriangColor}. 
The output of the adapted Restriction Decoder gives an appropriate correction for the triangular color code.

We conclude this section by mentioning that hook errors discussed in Ref.~\cite{DKLP02} form of subset of the space-time correlated errors leading to diagonal edges mentioned above, and discussed further in \cref{app:EdgeWeightCalc}. In our work, space-time correlated errors comprise of any error arising from a two-qubit gate failure which can have both spatial and/or temporal correlations. In addition, the flag qubits are used to detect and identify a subset of space-time correlated errors as discussed in \cref{sec:FTColorCodeCircuits}.

\section{Fault-tolerant implementation of the triangular color code}
\label{sec:FTColorCodeCircuits}
In \cref{subsec:CircuitLayout}, we first describe the stabilizer measurement circuits used for the triangular color code. The circuits are chosen to minimize the degree of the connectivity of each qubit. In \cref{subsec:FlagQubitInfo,subsub:FlagEdges,subsubEdgeRen}, we describe how flag qubits can be used to maintain the full effective distance of the Restriction Decoder in the presence of circuit level noise. In \cref{subsubsec:AltFlag}, we describe an alternative flag scheme compared to the scheme described in previous sections. 

\subsection{Triangular color code stabilizer measurement circuits}
\label{subsec:CircuitLayout}

\begin{figure*}
\centering
\includegraphics[width=0.8\textwidth]{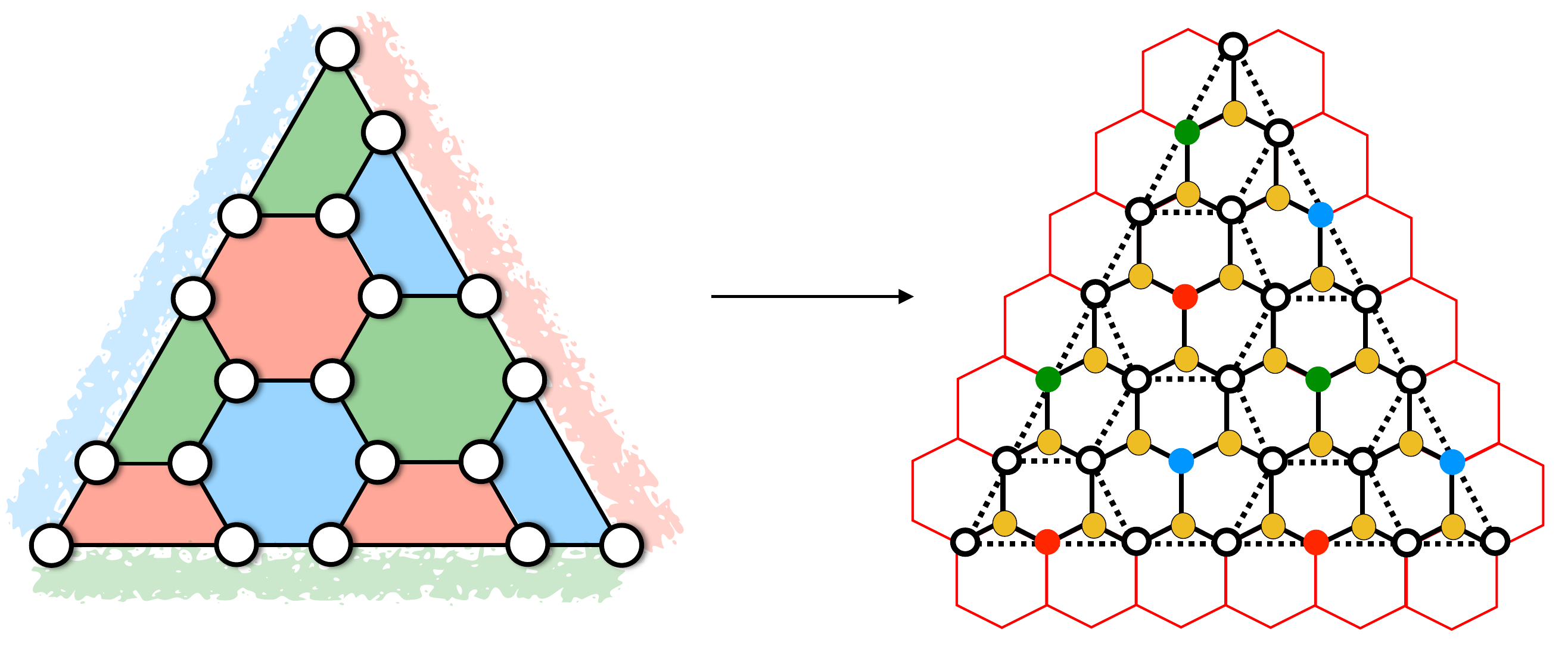}
\caption{Implementation of the triangular color code where syndrome measurement qubits (red, blue and green circles), flag qubits (yellow circles) and data qubits (white circles) have all degree three connectivity. All the qubits can be viewed as being located at the vertices of smaller hexagons (shown in red). The black edges between vertices represent physical connections between the qubits. The dashed edges do not represent any physical connections, and are simply included to highlight the two-dimensional lattice $\mathcal{L}$.}
\label{fig:LowDegColorCode}
\end{figure*}

As was shown in Ref.~\cite{CZYHC19}, for superconducting qubit architectures with fixed-frequency transmon qubits coupled via the cross resonance (CR) gates \cite{Rigetti2010,Chow2011}, reducing the degree of the connectivity between ancilla\footnote{In what follows, ancilla qubits will refer to both syndrome measurement and flag qubits.} and data qubits can minimize frequency collisions and reduce crosstalk errors. This motivates finding an implementation of the triangular color code where both data and ancilla qubits have low degree connectivity. 

\begin{figure*}
	\centering
	(a)\includegraphics[width=0.37\textwidth]{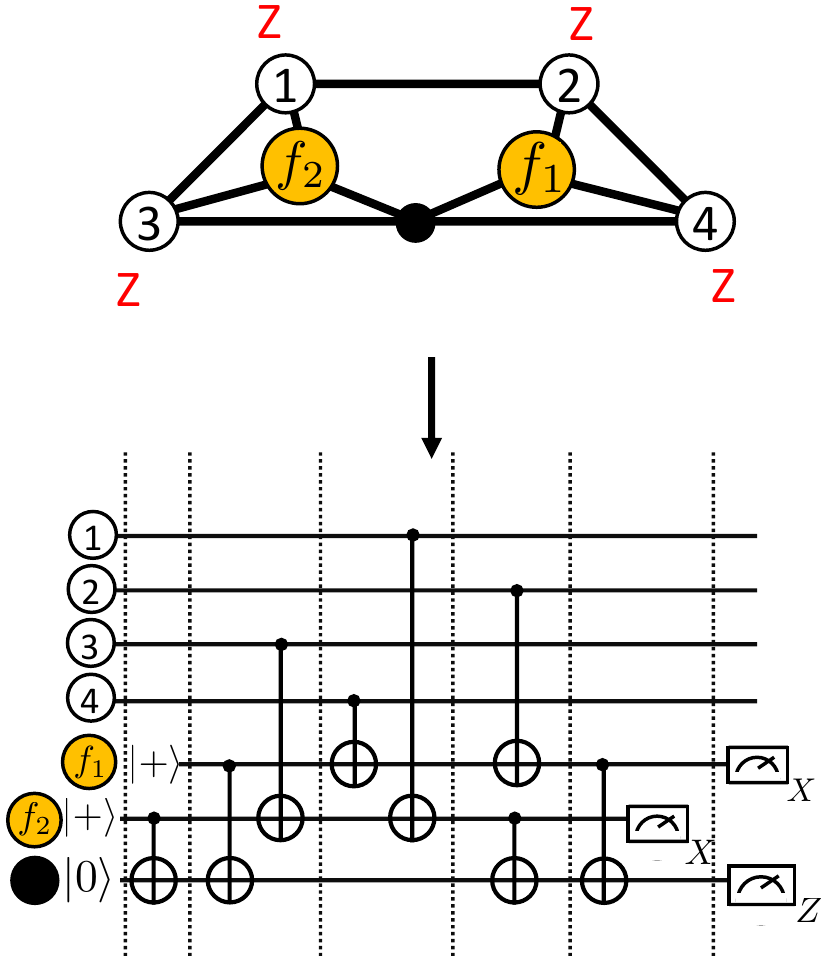}\hspace*{15mm}
        (b)\includegraphics[width=0.37\textwidth]{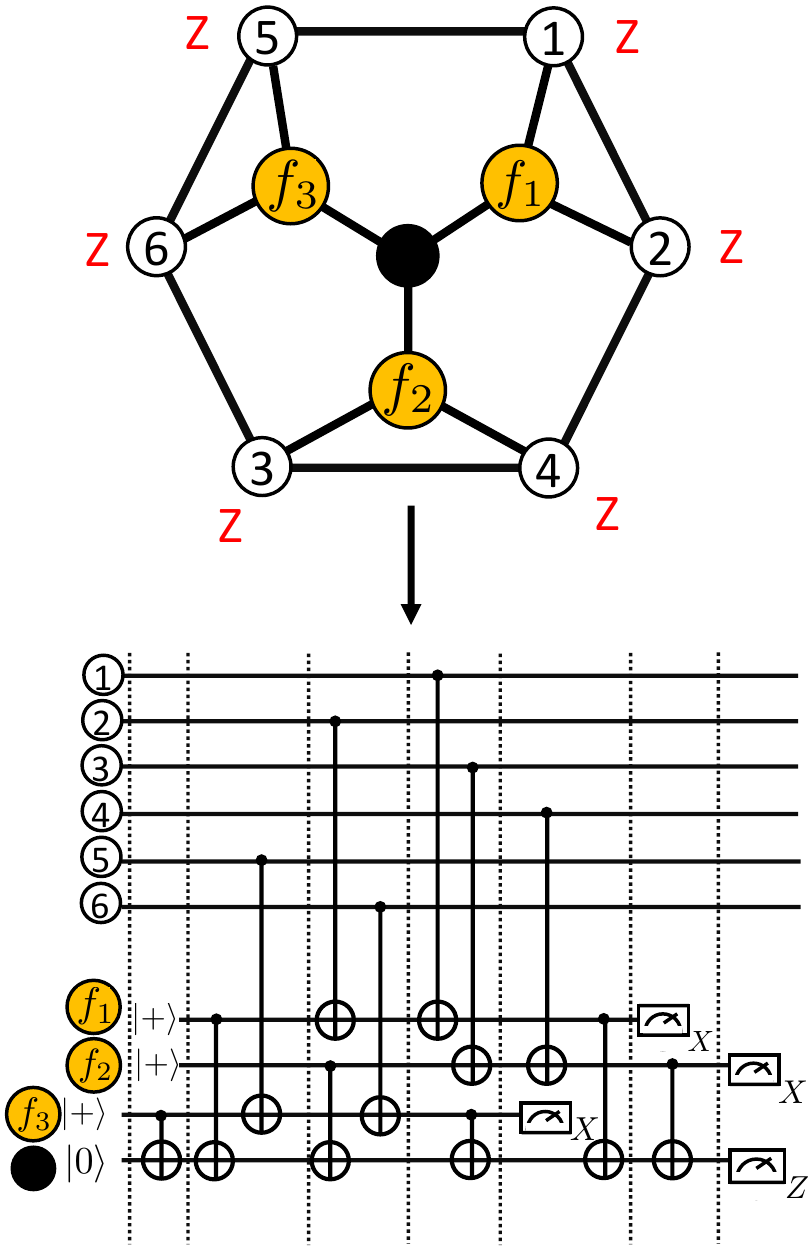}
	\caption{Circuits for measuring (a) weight-four and (b) weight-six $Z$-stabilizers of the triangular color code with minimal depth. The white circles correspond to data qubits, yellow circles to flag qubits and the dark circle to the ancilla qubits. Circuits for measuring the $X$-stabilizers can be obtained by reversing the direction of each CNOT gate,  swapping $\ket{+}$ and $\ket{0}$ states and swapping $X$ and $Z$ measurements.}
	\label{fig:StabCircMeas}
\end{figure*}

\begin{figure}
	\centering
	\includegraphics[width=0.5\textwidth]{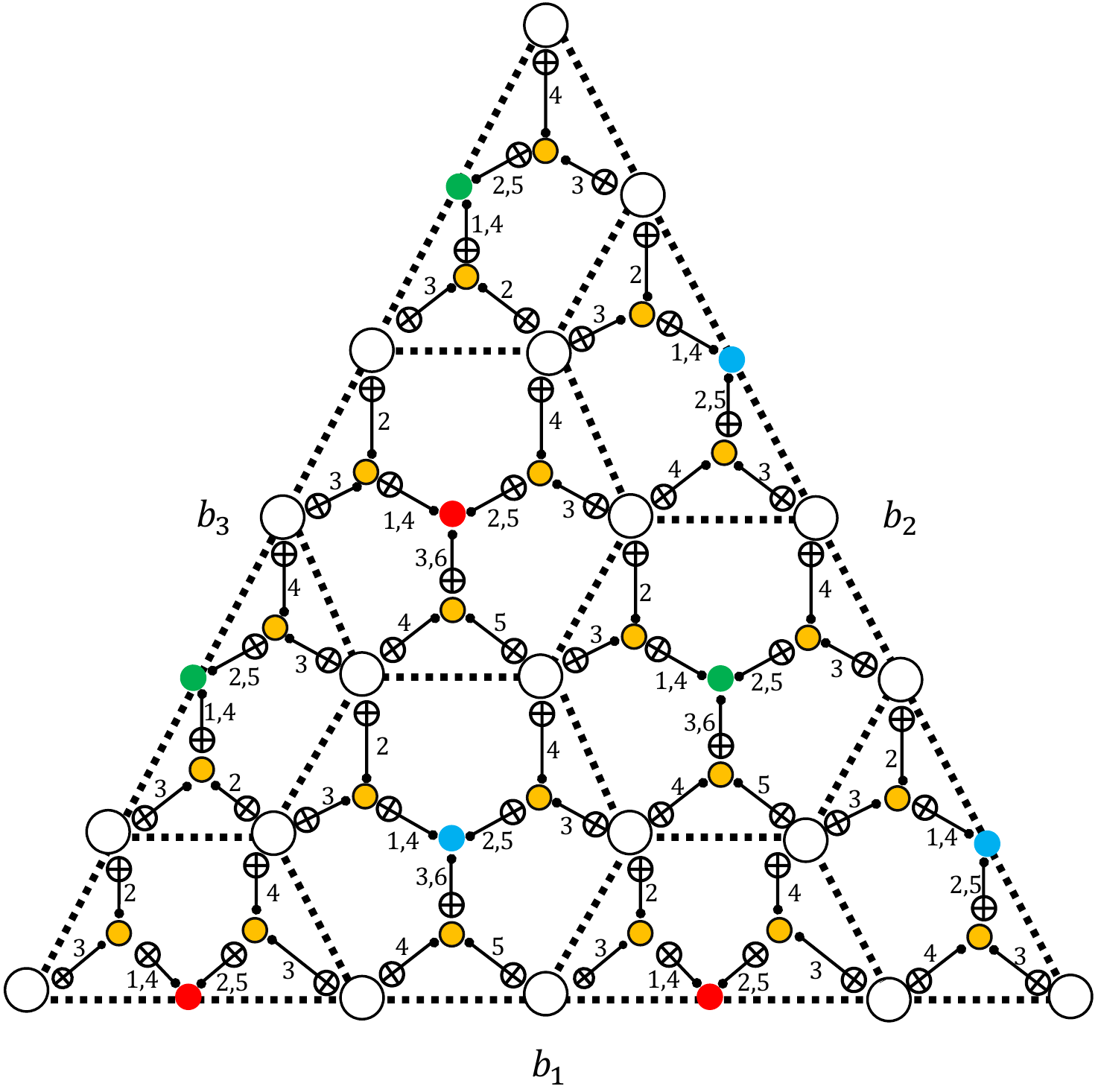}
	\caption{Full CNOT gate scheduling for one round of $X$ stabilizer measurements. Note that in order to minimize the total number of time steps, a different scheduling for the weight-four stabilizer measurements is used at the boundaries $b_1$, $b_2$ and $b_3$.}
	\label{fig:FullScheduling}
\end{figure}

In \cref{fig:LowDegColorCode}, an implementation of the triangular color code where both ancilla and data qubits have degree three connectivity is shown. Notice that both data and ancilla qubits arise from a tiling of smaller hexagons (shown in red). Since only one ancilla qubit is needed to measure a given stabilizer, the extra ancilla qubits can be used as flag qubits to correct high weight errors arising from fewer faults. Circuits for measuring the weight-four and weight-six stabilizers with minimal depth are provided in \cref{fig:StabCircMeas}. As can be seen, one round of $X$ stabilizer measurements can be done in 8 time steps (and thus 16 time steps are required to measure both $X$ and $Z$ stabilizers). For a distance $d \ge 5$ color code, the total number of data, syndrome measurement and flag qubits in the implementation of the code is $\frac{(3d-1)^2}{4}$.

The full CNOT scheduling for one round of $X$ stabilizer measurements which minimizes the total circuit depth is given in \cref{fig:FullScheduling}. If the same CNOT scheduling for the weight-four stabilizer measurements were used at the boundaries $b_1$, $b_2$ and $b_3$, an additional time step would be required to perform the $X$ stabilizer measurements. Hence a different scheduling for the weight-four stabilizers is used for each boundary. 

\subsection{Use of flag qubits to correct high weight errors}
\label{subsec:FlagQubitInfo}

\begin{figure}
	\centering
	\includegraphics[width=0.5\textwidth]{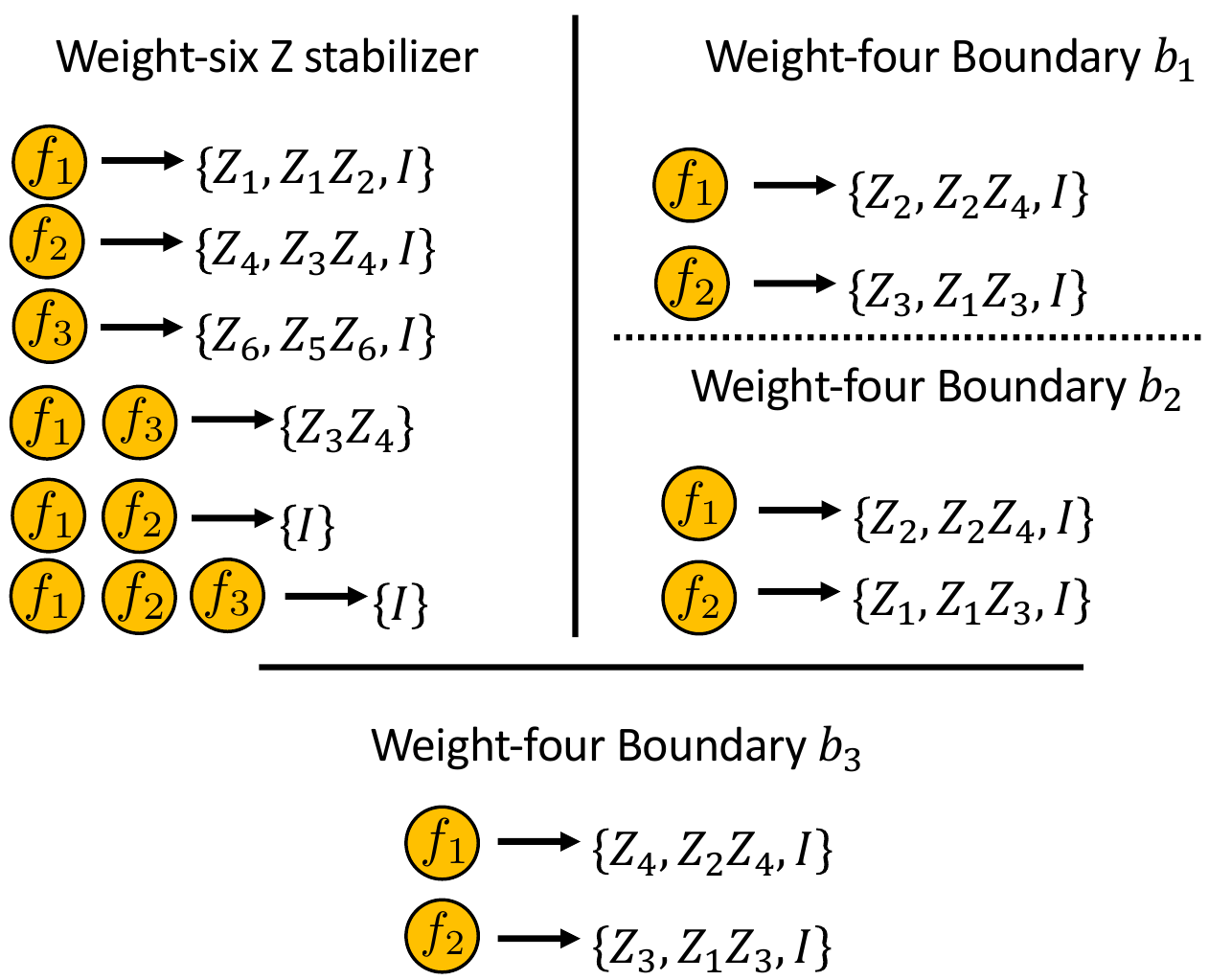}
	\caption{All possible $Z$-type data qubit errors arising from a single fault during a weight-six or weight-four $Z$ stabilizer measurement circuit resulting in non-trivial flag measurement outcome (see \cref{fig:StabCircMeas} for the placement of the flag qubits $f_1$, $f_2$ and $f_3$). For example, if a single fault results in only the flag qubit $f_1$ flagging, then the possible data qubit errors are $\{ Z_1, Z_1Z_2, I \}$. Note that we ignore contributions from CNOT failures which also add $X$ errors to the data since $X$ and $Z$ errors are decoded separately, and for a single fault, the weight of the $X$ errors can be at most one. Lastly, data qubit errors arising from a single fault during $X$-stabilizer measurements have the same support as shown above, but are of $X$-type.}
	\label{fig:OneFlagErrors}
\end{figure}

Performing an exhaustive search over all single fault locations in the circuits of \cref{fig:StabCircMeas}, it can be shown that a single fault results in a data qubit error of weight at most two. Similarly, two faults occurring during a weight-six stabilizer measurement can result in a data qubit error of weight at most three. In what follows, if a flag qubit has a non-trivial measurement outcome, we will say that the flag qubit flagged.

In order to maintain the full effective distance of the Restriction Decoder adapted to the triangular color code presented in \cref{sec:DecoSec}, Ref.~\cite{CB17} proves the sufficiency of 2-flag circuits for stabilizer measurements. In other words, if a single fault occurs in a circuit $C(g_i)$ for measuring a stabilizer $g_i$ which results in a data qubit error of weight greater than one, at least one flag qubit must flag\footnote{In general, a circuit for measuring a stabilizer P is called a t-flag circuit if at least one flag qubit flags whenever any $v \le t$ faults result in an error $E$ satisfying $\min{(\text{wt}(E), \text{wt}(EP ))} > v$.}. Similarly, if two faults occur in a circuit $C(g_i)$ for measuring a stabilizer $g_i$ resulting in a data qubit error of weight greater than two, at least one flag qubit must flag. By performing an exhaustive search, we find that the circuits in \cref{fig:StabCircMeas} are indeed 2-flag circuits. However, only 1-flag circuits are required to maintain the effective distance of a given decoder (not necessarily the Restriction Decoder) adapted the triangular color code when all circuit components can fail (see for instance the circuit level noise model in \cref{subsubEdgeRen}). We state this result as a theorem:

\begin{theorem}\label{thm:1flag_suff}
The triangular color code can be decoded with full distance if stabilizers are measured with 1-flag circuits.
\end{theorem}
The proof of \cref{thm:1flag_suff} is provided in \cref{app:more_thm}. To be clear, \cref{thm:1flag_suff} indicates that with full circuit level noise (where all components of the circuits are allowed to fail), one only needs to consider flag outcomes from a single fault in the stabilizer measurement circuits, which significantly simplifies the decoding scheme presented in \cref{subsub:FlagEdges,subsubEdgeRen}. However, to achieve the effective distance of a given decoder, one still needs to provide a flag based scheme which indicates how flag information can be used to recover the full effective distance. Such details are provided in \cref{subsub:FlagEdges,subsubEdgeRen}. Moreover, the theorem does not guarantee that there exists an efficient decoder achieving the full code distance. 

In \cref{fig:OneFlagErrors}, we give all possible data qubit errors arising from a single fault leading to non-trivial flag-qubit measurement outcomes. For weight-six $Z$-stabilizers, the only $Z$-type non-trivial data qubit error that can arise from a single fault resulting in two non-trivial flag measurements is $Z_{3}Z_{4}$ (where the flag qubits $f_1$ and $f_3$ have non-trivial measurement outcomes). Other errors arising from a single fault which results in more than one non-trivial flag measurement outcome cannot propagate to the data qubits. 

For weight-four stabilizers, since the CNOT scheduling is different at the three boundaries $b_1$, $b_2$ and $b_3$, the possible data qubit errors arising from a non-trivial flag measurement depends on the particular boundary and these features must be taken into account by the decoder\footnote{In particular, in how edge weights are assigned to edges of the lattice $\dual$ at the boundary.}. Lastly, note that a single fault occurring in a weight-four stabilizer measurement circuit can result in at most one non-trivial flag qubit measurement outcome. 

In order to correct weight-two errors arising from a single fault, we use similar methods to those presented in Ref.~\cite{CZYHC19}. There are two main steps that need to be implemented. First, edges corresponding to non-trivial flag measurement outcomes need to be added to the lattice $\dual$ described in \cref{sec:DecoSec} (such edges are given infinite weight unless a flag qubit associated with such an edge flags). Second, the weights for edges belonging to $\dual$ need to be renormalized, where the weights are chosen based on the number of flags and locations where they occur. 

\begin{figure*}
	\centering
	(a)\includegraphics[width=0.47\textwidth]{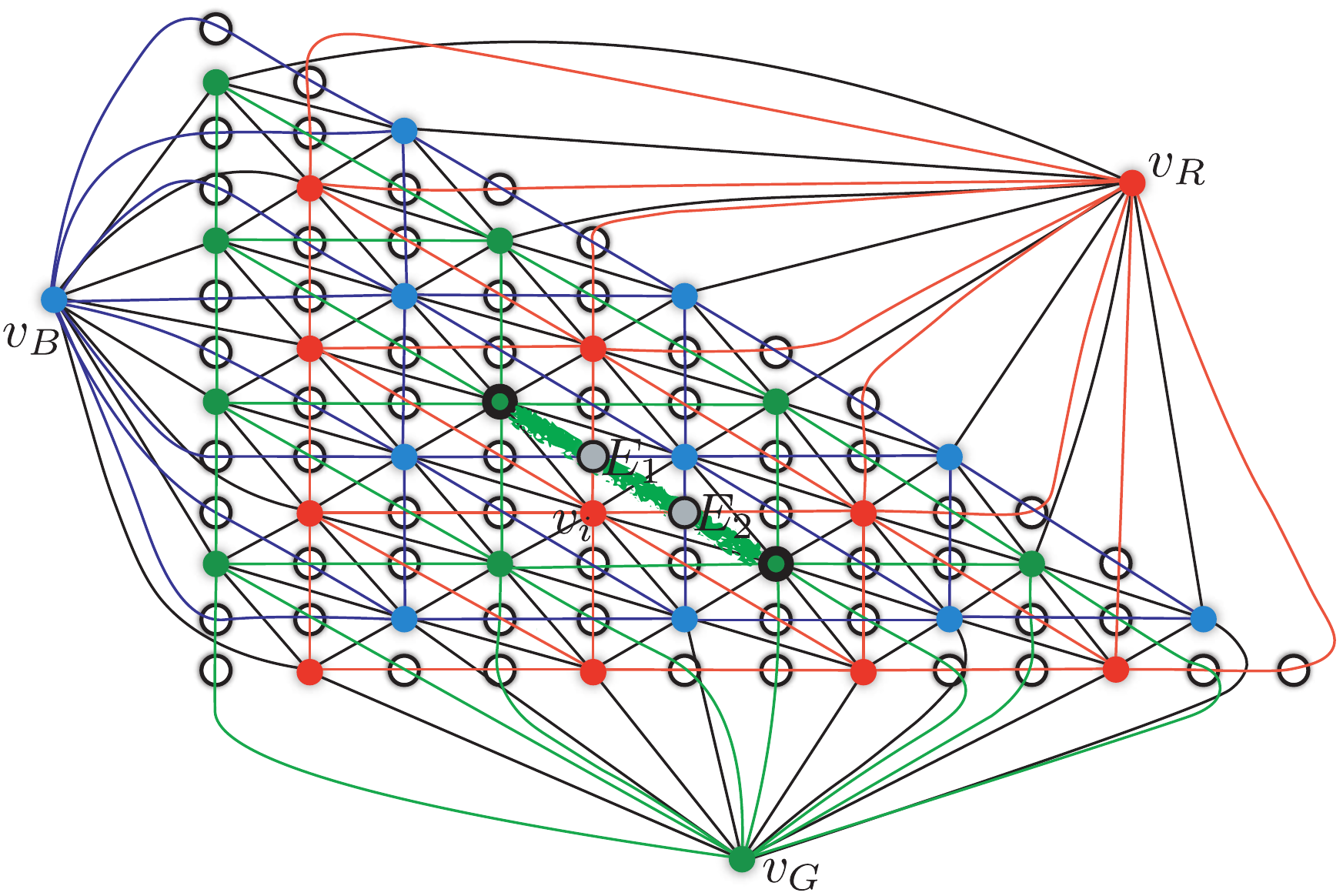}
        (b)\includegraphics[width=0.47\textwidth]{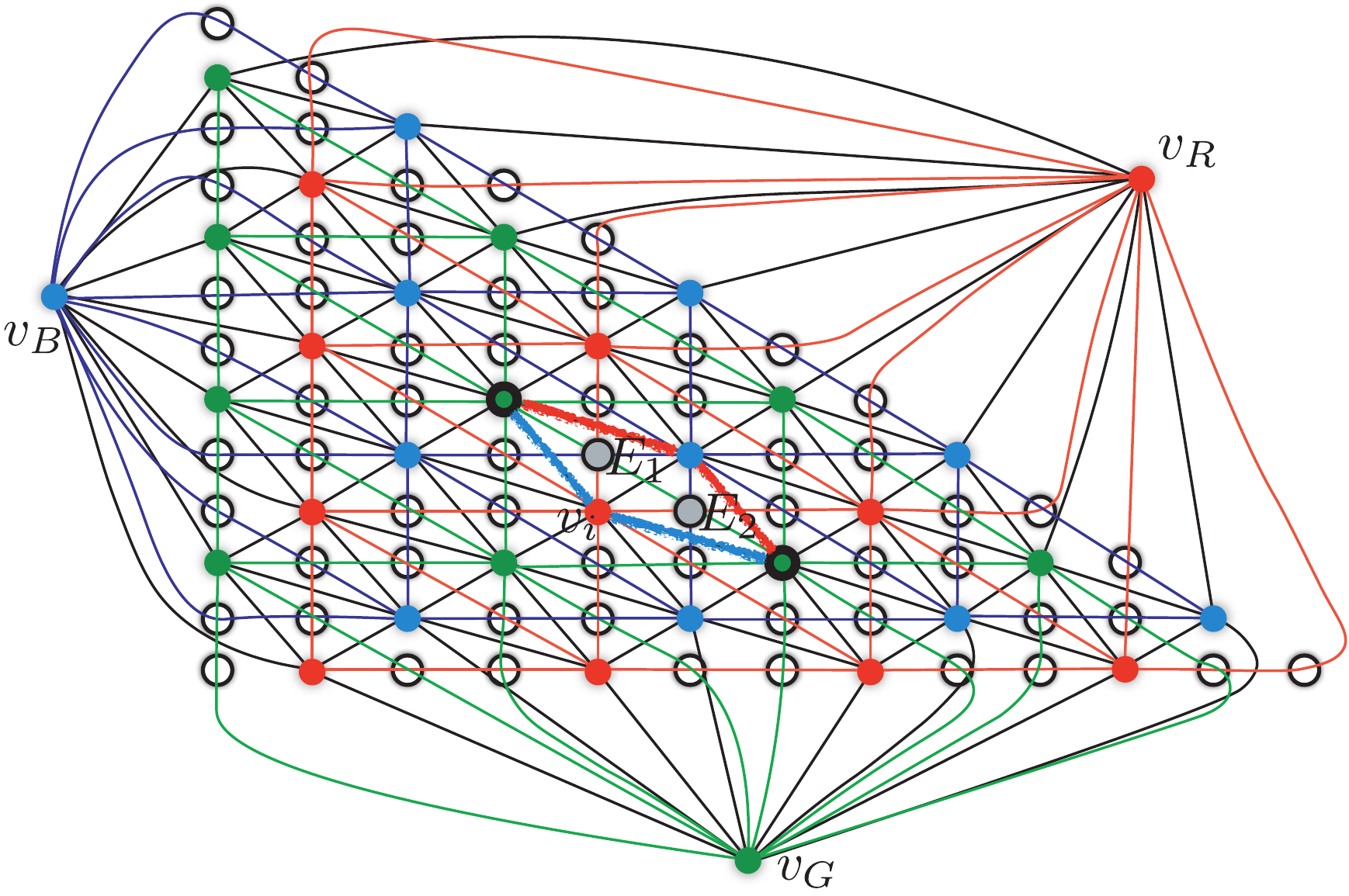}	
	\caption{Flag edges added to the 2D lattice $\dual$. In this example, the lattice $\dual$ shown corresponds to the $d=9$ triangular color code where the flag $f_1$ flagged (belonging to the face $i$ with ancilla vertex $v_i$), resulting in the data qubit errors $E_1E_2$. Note that this depiction of the lattice $\dual$  differs from previous illustrations, e.g., \cref{fig_stuff}, by an affine transformation.
	}
	\label{fig:FlagEdges}
\end{figure*}

\subsection{Constructing the matching graph $\mathcal{G}$}
\label{subsub:FlagEdges}

We begin by describing the particular flag edges that are added to $\face 1 \dual$ for one round of stabilizer measurements. Since a single fault causing a flag can result in a data qubit error of weight at most two, flag edges need to be added such that choosing such an edge during the MWPM step of the decoding algorithm would allow both data qubits to be identified when implementing the Restriction Decoder. In \cref{fig:FlagEdges}, we illustrate the 2D version of $\dual$ with the added flag edges, which connect two vertices in $\face 0 \dual$ of the same color, using the results from \cref{fig:OneFlagErrors}. For a weight-six stabilizer corresponding to a face $i$ of $\mathcal{L}$ with data qubits labelled 1 to 6 (as in \cref{fig:OneFlagErrors}), the possible weight two data qubit errors are arising from a single fault are $E_1E_2$, $E_3E_4$ and $E_5E_6$. Here $E_j$ is $X_j$ or $Z_j$ (depending on whether an $X$ or $Z$ type stabilizer is being measured) and has support on the data qubit $j$ belonging to the face $i$ of $\mathcal{L}$. Since such weight two-data qubit errors results in two highlighted vertices of the same color\footnote{Our convention is that a flag edge connecting two vertices of the same color will have the same color as the two vertices.}, the vertex $v_i \in \face 0 \dual$ corresponding to the face $i$ should encircled by three flag edges connecting vertices in $\face 0 \dual$ of the same color. The flag edges are chosen to overlap with data qubits $\{1, 2 \}$, $\{3, 4 \}$ and $\{5, 6 \}$.

As an example, in \cref{fig:FlagEdges} (a), we illustrate a highlighted green flag edge arising from a weight-two data qubit error $E_1E_2$ connecting two highlighted green vertices in $\face 0 \dual$. In other words, we are considering the case where a single fault during a weight-six stabilizer measurement circuit corresponding to face $i$ with red vertex $v_i \in \face 0 \dual$ resulted in the weight-two data qubit error $E_1E_2$. The weight of the flag edge, along with all other edges in $\face 1 \dual$, are then renormalized (see \cref{subsubEdgeRen} for a description of the renormalization step). If there are no other faults, then the green flag edge will be chosen during the MWPM step of the Restriction Decoder as illustrated in \cref{fig:FlagEdges} (a). On the other hand, if the two data qubit errors $E_1E_2$ had arisen from failures at the qubits 1 and 2, then the four edges shown in \cref{fig:FlagEdges} (b) would be highlighted during the MWPM step. Hence, prior to implementing the local lifting procedure \lift\ of the Restriction Decoder described in \cref{sec:DecoSec}, the highlighted flag edge of \cref{fig:FlagEdges} (a) is projected to the four 2D edges of $\face 1 \dual$ shown in \cref{fig:FlagEdges} (b) (again, assuming that a single fault occurred resulting in the flag $f_1$).

Note that flag edges are used specifically for weight-two errors arising from a single fault (we will discuss more about weight one data qubit errors arising from a single fault resulting in a flag in \cref{subsubEdgeRen}). In general, a given flag edge can be chosen in at most two restricted lattices when perform MWPM. In the example illustrated by \cref{fig:FlagEdges}, the restricted lattices are $\dualx{RB}$ and $\dualx{GB}$. If a flag edge is highlighted when performing MWPM on a given restricted lattice (say $\dualx{RB}$), the flag edge should be projected to two 2D edges belonging to $\Delta_1 ( \dualx{RB})$ (i.e. edges present in $\Delta_1 ( \dualx{RB})$ before flag edges were introduced). In the presence of other errors, it is possible that a flag edge would only be chosen during MWPM in only one of the two restricted lattices it belongs to, in which case one would project the flag edge onto \textit{two} 2D edges of $\face 1 \dual$ instead of four. Lastly, note that two-qubit errors arising when there are flags (i.e. those given in \cref{fig:OneFlagErrors}) always result in two highlighted vertices of the same color (assuming there are no other errors). This argument justifies our choice of flag edges always connecting two vertices of the same color.

\subsection{Edge weight renormalization}
\label{subsubEdgeRen}

\begin{figure*}
	\centering
	(a)\includegraphics[width=0.47\textwidth]{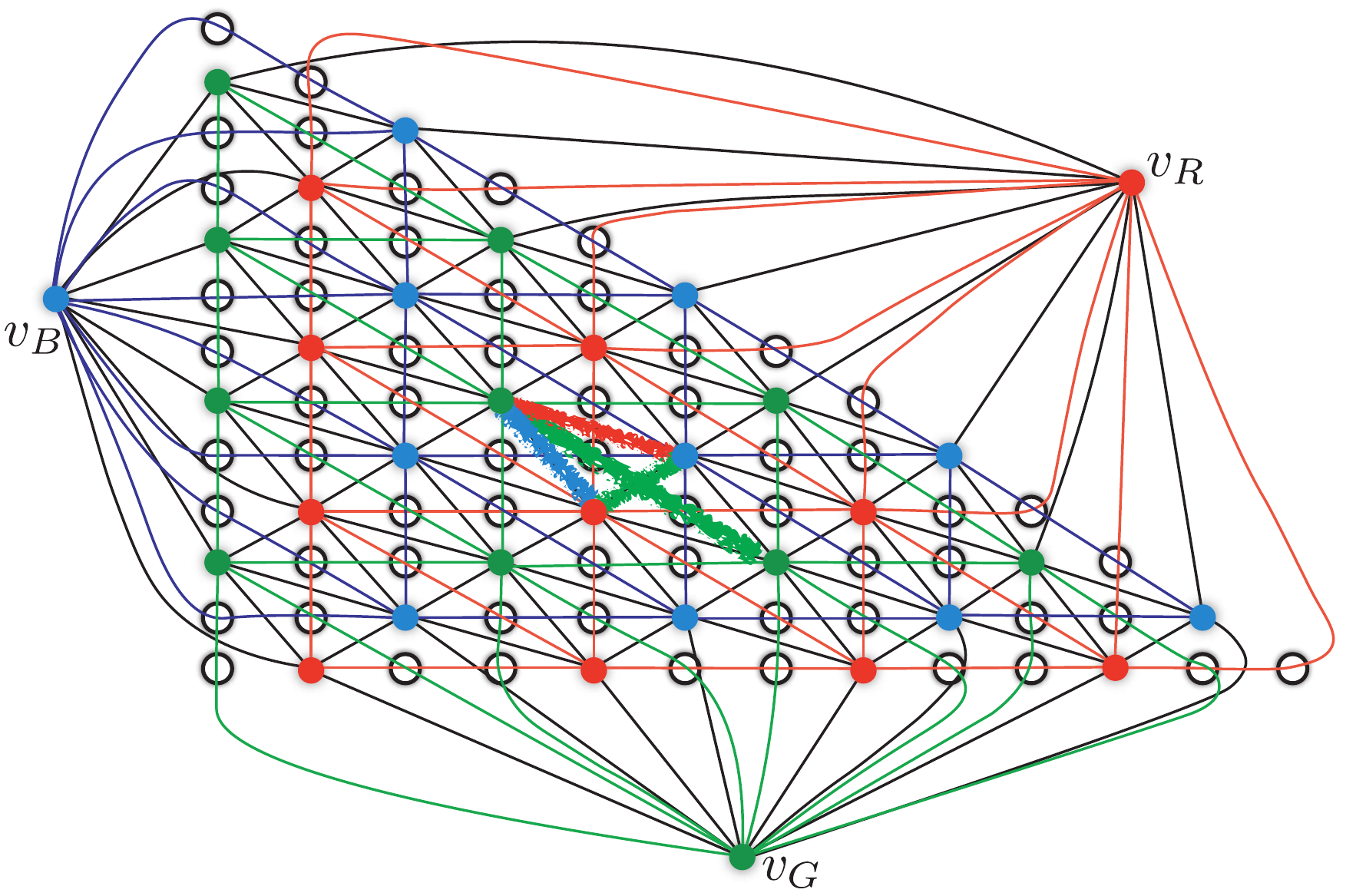}
        (b)\includegraphics[width=0.47\textwidth]{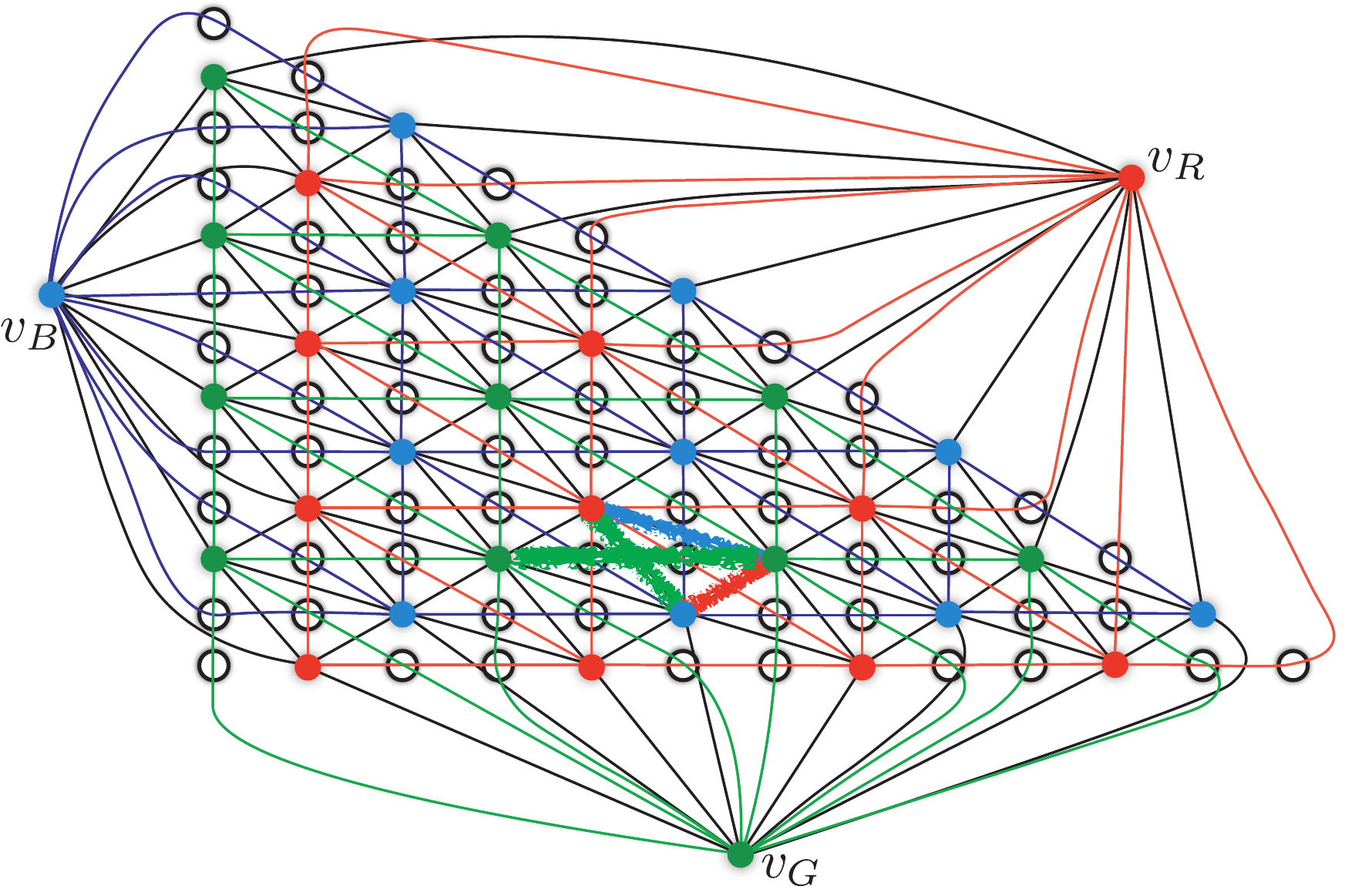}	\\
         (c)\includegraphics[width=0.47\textwidth]{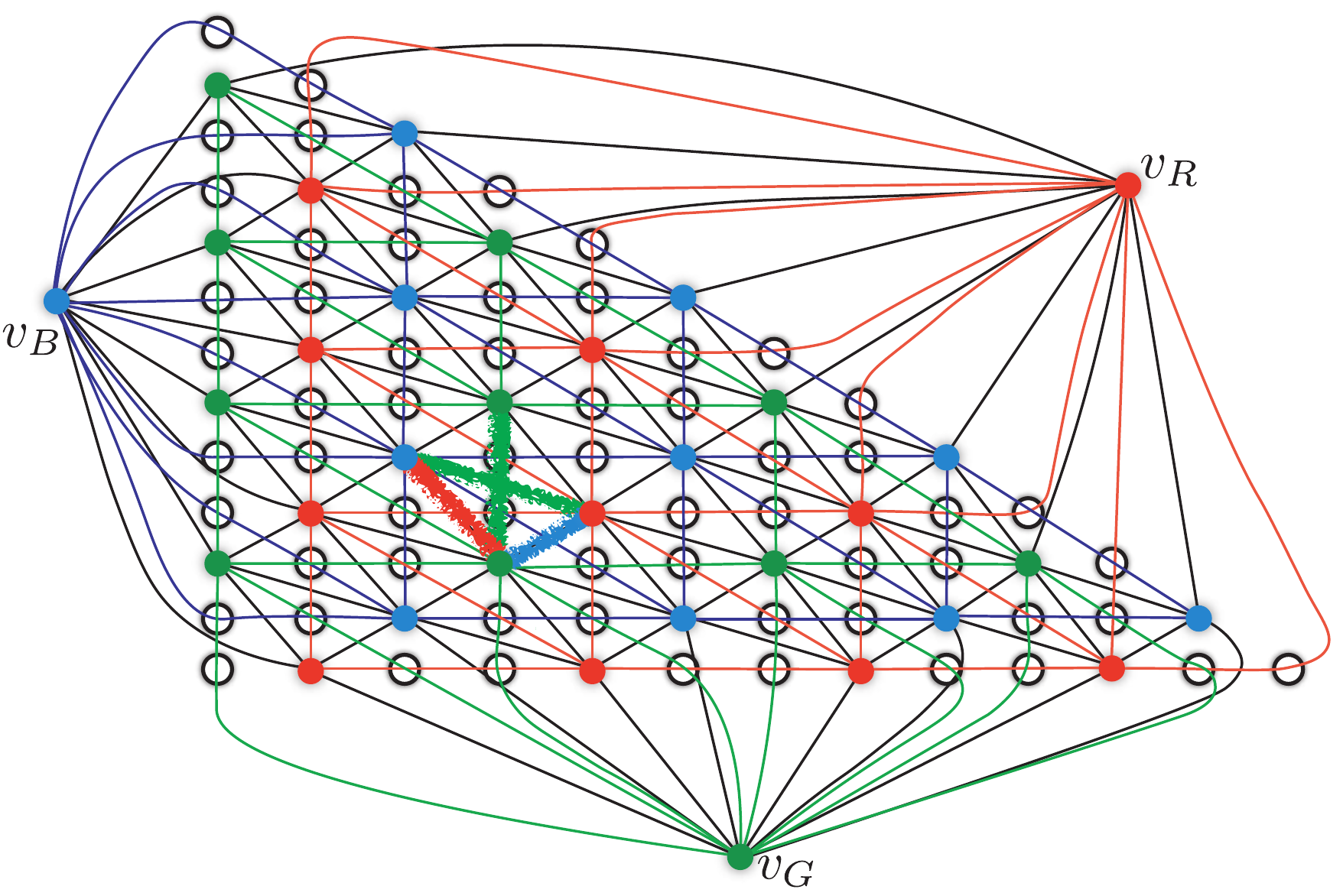}	
	\caption{Flag edge and 2D edges belonging to $\Delta_1 (\dual)$ associated with the flag outcomes (a) $f_1$, (b) $f_2$ and (c) $f_3$ for a given face $i$ in $\mathcal{L}$. As can be seen from \cref{fig_stuff}, a weight-one data qubit error results in three highlighted edges in $\Delta_1 (\dual)$. Hence the flag edges, in addition to edges for the possible weight-one data qubit errors, form the edges shaped as an arrow as seen in the figure and should be renormalized to edge weights with error probability $\mathcal{O}(p)$ in the presence of the corresponding flags. The same pattern is chosen for blue and red flag edges. If both flag qubits $f_1$ and $f_3$ flag (which can only result in a weight-two data qubit error $E_1E_2$), then only the flag edge in (b) should be renormalized to an edge weight with error probability $\mathcal{O}(p)$.}
	\label{fig:AllRenormFlags}
\end{figure*}

 The circuit level depolarizing noise model used throughout the manuscript is given as follows:

\begin{enumerate}
        \item With probability $p$, each single-qubit gate location is followed by a Pauli error drawn uniformly and independently from $\{ X,Y,Z \}$.
	\item With probability $p$, each two-qubit gate is followed by a two-qubit Pauli error drawn uniformly and independently from $\{I,X,Y,Z\}^{\otimes 2}\setminus \{I\otimes I\}$.
	\item With probability $\frac{2p}{3}$, the preparation of the $\ket{0}$ state is replaced by $\ket{1}=X\ket{0}$. Similarly, with probability $\frac{2p}{3}$, the preparation of the $\ket{+}$ state is replaced by $\ket{-}=Z\ket{+}$.
	\item With probability $\frac{2p}{3}$, any single qubit measurement has its outcome flipped.
	\item Lastly, with probability $p$, each idle gate location is followed by a Pauli error drawn uniformly and independently from $\{ X,Y,Z \}$.
\end{enumerate}

Let $P_{e}$ be the probability for a given edge $e \in \face 1 \dual$ to be highlighted during MWPM. $P_{e}$ can be computed by summing the probabilities of all error configurations (using the circuit level depolarizing noise model described above) resulting in the edge $e$ being highlighted. The weight for the edge $e$ is then given by $w_{e} = -\log{P_{e}}$. More details on the edge weight calculations for the lattices $\dual$ considered in this manuscript are given in \cref{app:EdgeWeightCalc}.

Let $\mathcal{S} = \langle g_1, g_2, \cdots, g_r \rangle$ be the generating set for the stabilizer group of the triangular color code. Further, let $n_{g_{i}} = 1$ if  there are flags corresponding to the configuration in \cref{fig:OneFlagErrors} during the measurement of $g_i$, and $n_{g_{i}} = 0$ otherwise. Now consider the case where $m > 0$ stabilizers flagged, i.e. $m = \sum_{g_{i} \in \mathcal{S}} n_{g_{i}}$ with at least one $n_{g_i}$ being non-zero. Any other error arising from faults which don't cause any flags must occur with probability $\mathcal{O}(p^{m+l})$ where $l \ge 1$. in Ref.~\cite{CZYHC19}, it was shown that in such a case, all edges $e$ in the matching graphs that cannot contain errors resulting from the set of $m$ flags (with error probabilities $P_{e}$) should be renormalized by $P'_{e} = p^{m}P_{e}$, whereas edges $e_{f}$ that could contain errors resulting from the flags should have edge weights $w_{e_{f}} = -\log{P_{e_{f}}}$ with error $P_{e_{f}} = \mathcal{O}(p)$, which is computed by considering all single faults leading to the particular flag outcome (see \cref{app:EdgeWeightCalc}). Further it was shown that by adopting such a scheme, the full distance of the considered codes could be preserved (i.e. any error arising from at most $\floor*{(d-1)/2}$ faults would be corrected). 
In this work, edge weights will be renormalized as described above allowing any set of errors from at most $\sim d/3$ faults to be corrected (since the Restriction Decoder can correct errors arising from at most $\sim d/3$ faults).

Since a flag edge $e_{f_i}$ should only be used when there are flags, its weight should be set to infinity unless the flag qubits associated with the edge  $e_{f_i}$ flagged. Further, as seen in \cref{fig:OneFlagErrors}, a single fault resulting in flags can also introduce weight-one data qubit errors. In \cref{fig:AllRenormFlags}, we show the flag edges and edges associated with single qubit errors whose weights should be renormalized to $w_{e_{f_i}} = -\log{P_{e_{f_i}}}$ with error $P_{e_{f_i}} = \mathcal{O}(p)$ for the possible flag outcomes of \cref{fig:OneFlagErrors}. Note that in \cref{fig:AllRenormFlags}, we considered green flag edges centered around a red ancilla vertex in $\Delta_0 (\dual)$. However the same pattern of flag edges would be chosen for blue flag edges centered around a green ancilla vertex in $\Delta_0 (\dual)$, and red flag edges centered around a blue ancilla vertex in $\Delta_0 (\dual)$. 

To summarize, suppose are $m = \sum_{g_{i} \in \mathcal{S}} n_{g_{i}}$ flags during the stabilizer measurements with $m>0$. Define $C_f$ to be the set of edges in $\Delta_1 (\dual)$ associated with the flag outcomes, in addition to the 2D edges in $\Delta_1 (\dual)$ associated with the possible single-qubit errors arising from faults resulting in the flags. Then all edge weights for edges $e_{i} \in C_f$ should be renormalized to $w_{e_{i}} = -\log{P_{e_{i}}}$ with $P_{e_{i}} = \mathcal{O}(p)$. $P_{e_i}$ is computed by considering all single fault locations and their associated probabilities leading to a flag with its corresponding edge. Further, the edge weights for all edges $e_j \notin C_f$ should be renormalized to  $w_{e_{j}} = -\log{P'_{e_{j}}}$ with $P'_{e_j} = p^{m}P_{e_j}$. If $e_j \notin C_f$ is a flag edge, then its weight should be infinite. 

\subsection{Alternative flag scheme}
\label{subsubsec:AltFlag}

\begin{figure}
	\centering
	\includegraphics[width=0.45\textwidth]{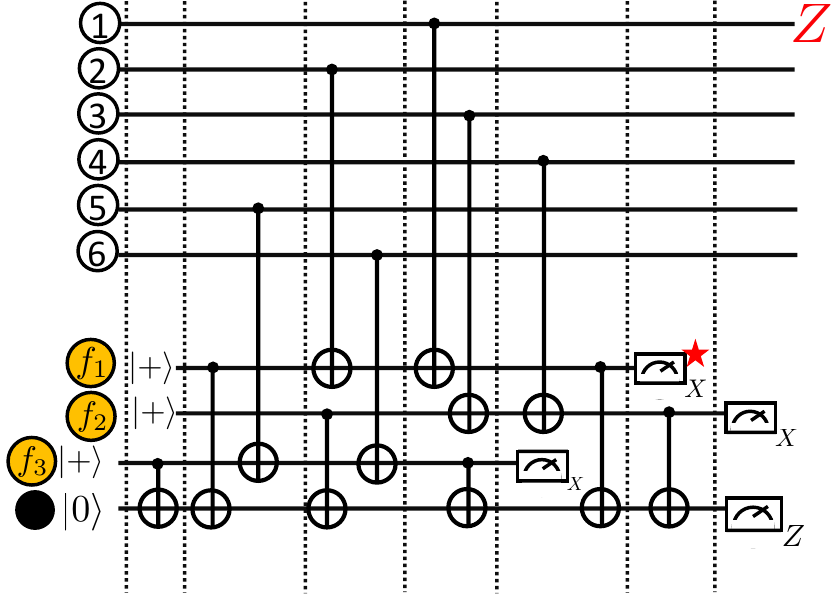}
	\caption{Suppose that a single fault occurs during the weight-six $Z$ stabilizer measurement resulting in the flag $f_1$. From \cref{fig:OneFlagErrors}, it can be seen that the possible $Z$-type data qubit errors are $\{ I, Z_1, Z_1Z_2 \}$. Applying the correction $Z_1$ immediately after the flag outcome guarantees that there can be at most a weight-one data qubit error.}
	\label{fig:DirectFlagScheme}
\end{figure}

Instead of renormalizing the edge weights for edges in $\Delta_1 (\dual)$ based on the flag measurement outcomes, given enough flag qubits associated with a stabilizer measurement circuit, there is a simpler approach that one can take\footnote{The ideas presented in what follows were first proposed in a correspondence with Ben Reichardt regarding the weight-four stabilizer measurements of the heavy-hexagon code of \cite{CZYHC19}.}.

Consider the the case where only the flag qubit $f_1$ flags during the weight-six $Z$ stabilizer measurement shown \cref{fig:DirectFlagScheme}. Assuming there was at most one fault, from \cref{fig:OneFlagErrors}, the possible $Z$-type data qubit errors are $\{ I, Z_1, Z_1Z_2 \}$. Consequently, if one applies the correction $Z_1$ to the data immediately after the flag outcome $f_1$ is known, the weight of any remaining data qubit error can be at most one. Similarly, for a flag outcome $f_2$, one would apply the correction $Z_4$, for $f_3$ one would apply the correction $Z_6$ and for $f_1f_3$, one would apply $Z_3Z_4$. Lastly, if a different flag outcome is obtained, no correction is applied to the data\footnote{To be clear, a correction to the data would still be applied once all the measurement outcomes were obtained and fed into the Restriction Decoder described in \cref{sec:DecoSec}. The corrections described in this section refer to preliminary corrections, based uniquely on the flag measurement outcomes.}. In all cases, the remaining data qubit errors arising from a single fault during the measurement of the stabilizer can be at most one. The same scheme can be applied when measuring $X$ stabilizers, but replacing the $Z$ corrections with $X$ Pauli's, supported on the same qubits. Further, one can define similar rules for the weight-four stabilizer measurements. Also, note that in the presence of slow measurements, such corrections based on the above flag outcomes could be done in a Pauli frame \cite{Knill05,DA07,Barbara15,CIP17}.

In what follows, the flag scheme described above will be referred to as the \textit{direct flag method}. By applying the direct flag method, it is straightforward to see that a single fault occurring during a stabilizer measurement can result in a data qubit error of weight at most one. From \cref{thm:1flag_suff}, if two faults occur resulting in a flag outcome compatible with those of \cref{fig:OneFlagErrors}, the resulting data qubit error can be of weight at most three and cannot be entirely supported on a logical operator. With these properties, it is straightforward to show that the direct flag method satisfies the the fault-tolerant definitions of \cite{CB17}.

One caveat of such a scheme is that in general, more flag qubits are required for each stabilizer measurement compared to the edge weight renormalization scheme of \cref{subsubEdgeRen}. For instance, consider the case where a single flag qubit was used for a weight-four $Z$-type stabilizer. If a single fault resulted in a flag, the possible $Z$-type data qubit errors would be $\{Z_1, Z_3, Z_1Z_2, Z_3Z_4,I \}$. Since there is only one flag qubit, one would not have enough information to determine whether to apply a $Z_1$ or $Z_3$ correction to the data. However, one could still use the scheme of \cref{subsub:FlagEdges,subsubEdgeRen} and renormalize the edge weights for edges in $\Delta_1 (\dual)$ corresponding to data qubit errors $Z_1$, $Z_3$ and $Z_1Z_2$ (using the same methods as in \cref{fig:AllRenormFlags}). Following the same arguments presented in Ref.~\cite{CZYHC19}, one can show that the effective code distance of the color code would be preserved. 

Another caveat of the direct flag method (which is relevant for the lattice of \cref{fig:FullScheduling}) is that regardless of the noise model, the same operations are always applied to the data (and thus the scheme is suboptimal). In the case where measurement errors occur with high probability, the direct flag method would apply weight-one corrections to the data more often than necessary, whereas the renormalization methods of \cref{subsub:FlagEdges,subsubEdgeRen} would incorporate the higher measurement error probabilities into the assignment of edge weights. 

For the reasons mentioned above, the numerical results of \cref{sec:NumRes} were obtained using the edge weight renormalization methods of \cref{subsub:FlagEdges,subsubEdgeRen}.

\section{Numerical results}
\label{sec:NumRes}

In this section we present the plots illustrating the logical failure rates for both code capacity noise and the full circuit level noise model described in \cref{subsubEdgeRen}. 

\begin{figure}
	\centering
	\includegraphics[width=0.46\textwidth]{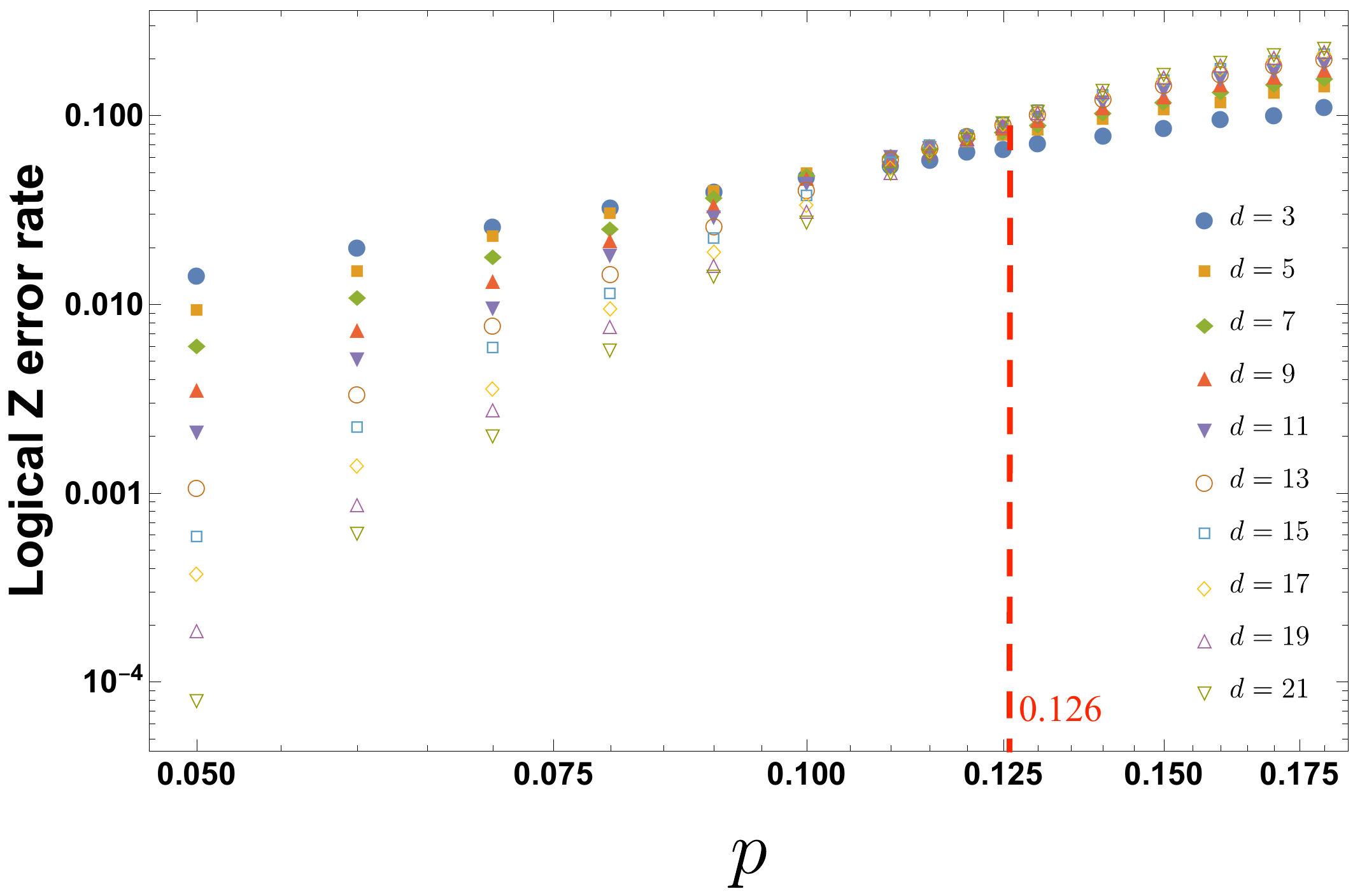}
	\caption{Logical $Z$ error rates for the triangular color code afflicted by code capacity noise, where each data qubit has an $X$, $Y$ or $Z$ Pauli error with probability $p/3$ each. Data qubit errors were corrected using the 2D decoder described in \cref{sec:DecoSec}. Since the decoder corrects both $X$ and $Z$ errors symmetrically, the plot for logical $X$ errors is identical to the one shown above.}
	\label{fig:2DLogZPlot}
\end{figure}

\begin{figure*}
	\centering
(a)\includegraphics[width=0.46\textwidth]{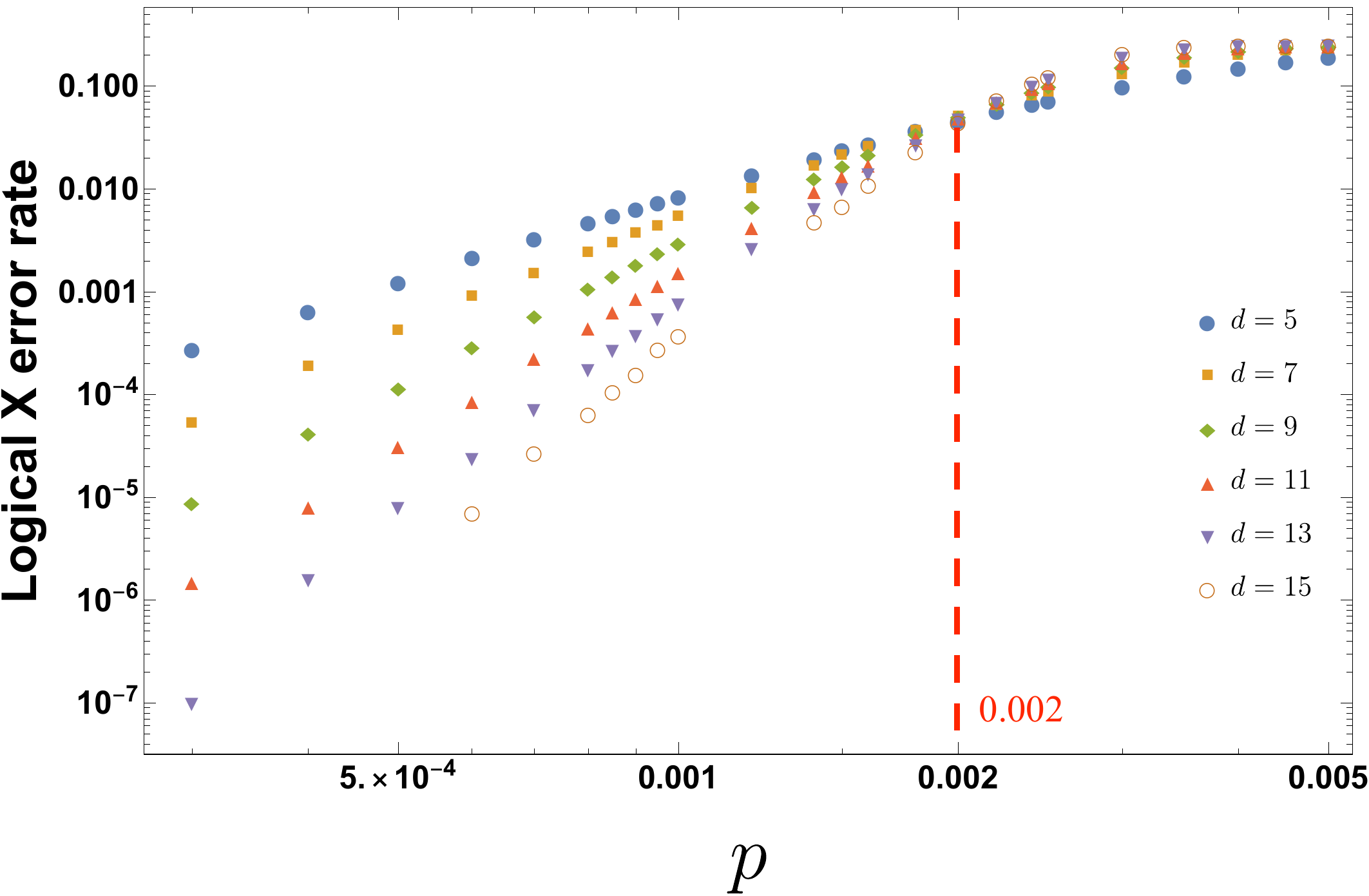}\hspace*{3mm}
(b)\includegraphics[width=0.46\textwidth]{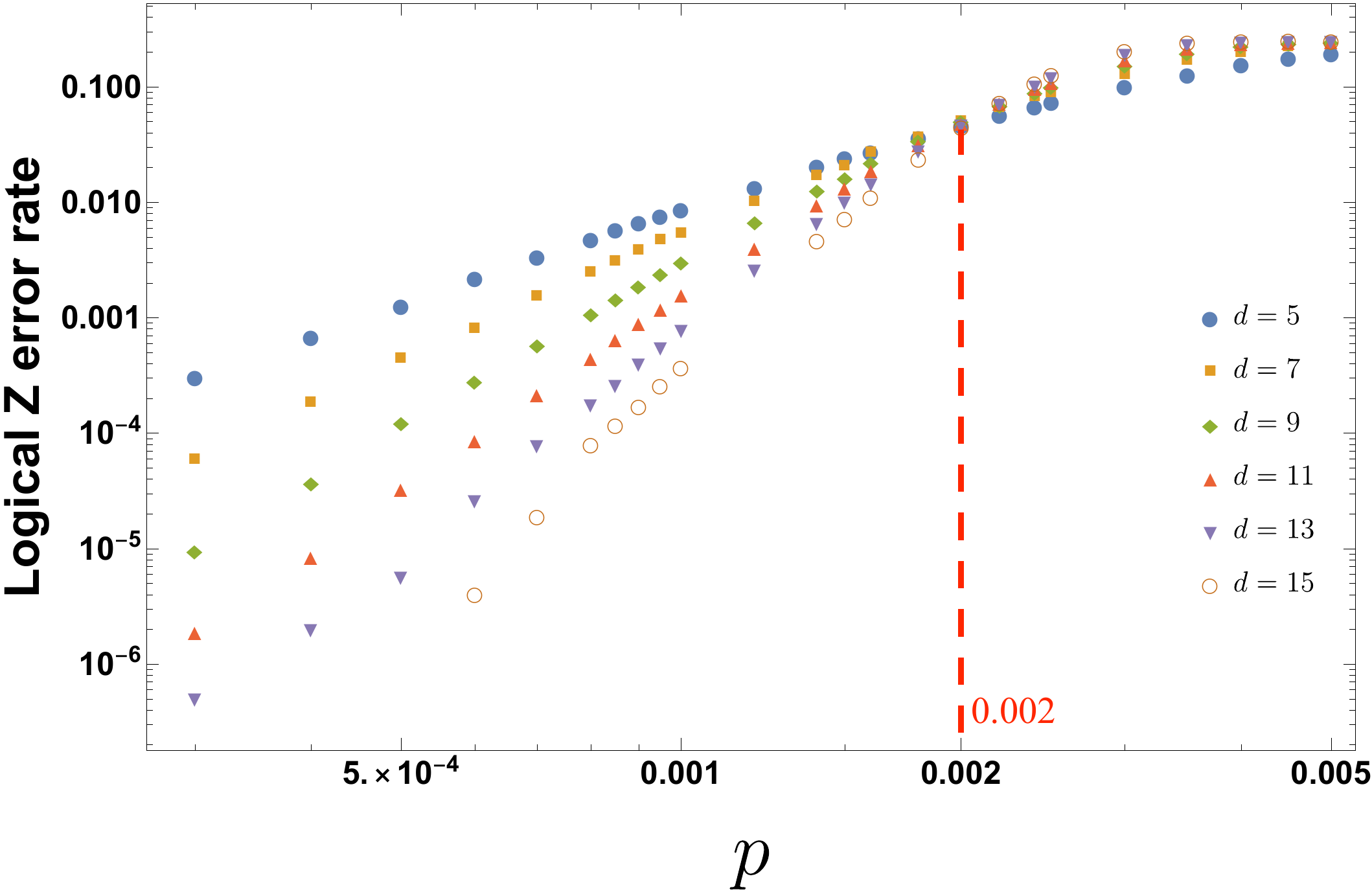}
	\caption{(a) Logical $X$ error rates and (b) logical $Z$ error rates for the triangular color code afflicted by the full circuit level noise model described in \cref{subsubEdgeRen}. For both logical $X$ and $Z$ errors, the threshold is found to be $p_{\text{th}} = 0.002$. }
	\label{fig:3DLogErr}
\end{figure*}

For code capacity noise, each data qubit can be afflicted by $X$, $Y$ or $Z$ Pauli errors, each occurring with probability $p/3$. Measurements, state preparation and gates are assumed to be implemented perfectly. Thresholds for code capacity noise are important as they illustrate the theoretical limitations of a code and the decoder used to correct errors. In \cref{fig:2DLogZPlot}, we illustrate logical $Z$ error rates for the triangular color code using the 2D decoder presented in \cref{sec:DecoSec}. We considered distances $d=3$ to $d=21$. Note that since both $X$ and $Z$ errors are corrected symmetrically by the decoder, the plot for logical $X$ error rates is identical to the one shown in \cref{fig:2DLogZPlot}. 

Numerically, we find a threshold of $12.6 \%$, in accordance to the threshold found in Ref.~\cite{MKJO19} using the decoder by projection. Although the threshold is identical to the decoder by projection, the Restriction Decoder applied to the triangular color code presented in \cref{sec:DecoSec} achieves lower logical error rates since a larger family of errors can be corrected by the latter (see \cref{sec_naive} for more details).

In \cref{fig:3DLogErr} we illustrate the logical $X$ and $Z$ error rates for the triangular color code afflicted by the full circuit level noise model using the Restriction Decoder described in \cref{Subsec:IncMeasErrs} (with $T = d + 1$, where no errors are introduced in round $d+1$ to guarantee projection to the codespace) along with the flag methods and scheme to renormalize edge weights described in \cref{sec:FTColorCodeCircuits,app:EdgeWeightCalc}. As can be seen, for both $X$ and $Z$ logical failure rates, the threshold occurs at $p_{\text{th}} = 0.002$.

Note that in our simulations, for a given syndrome measurement round, we chose the convention where we first measured $X$-stabilizers followed by $Z$-stabilizers. Now, suppose that during the $j$'th syndrome measurement round, a subset of flag qubits flagged during the $X$-stabilizer measurements. Flag edges (described in \cref{subsub:FlagEdges}) with finite weights are then introduced in the $j$'th 2D layer of the full 3D lattice $\dual$ used for $X$-stabilizer measurement outcomes. However, for flag qubits which flag during the $Z$-stabilizer measurements, $Z$ errors resulting from faults which led to the non-trivial flag measurements would only be detected during the $j+1$'th syndrome measurement round. Hence, in such a case, flag edges with finite weight must be introduced in the $j+1$'th 2D layer of the full 3D lattice $\dual$ used for $Z$-stabilizer measurement outcomes. We also point out that in order to obtain the high threshold and logical error rate scaling in \cref{fig:2DLogZPlot}, the choice of flag edges and structure of the decoding graphs given in \cref{subsub:FlagEdges}, and the edge weight renormalization scheme given in \cref{subsubEdgeRen} were all crucial features. As shown in \cite{CZYHC19}, omitting such steps can lead to higher logical error rates by several orders of magnitude. 

Lastly in \cref{subsubEdgeRen}, we showed that in the presence of flags, edges $e$ are renormalized as $w_e = -\log{P'_e}$ with $P'_e = p^{m}P_e$ where $m = \sum_{g_{i} \in \mathcal{S}} n_{g_{i}}$ corresponds to the number of stabilizer measurement circuits which flagged. We numerically explored a more general setting, where $P'_e = p^{\alpha m}P_e$ for some parameter $\alpha$. Choosing $\alpha = 1.5$ (compared to $\alpha = 1$ in previous simulations), we did not find a significant difference between the obtained logical failure rate curves and those of \cref{fig:3DLogErr}. We leave a more detailed exploration of renormalization parameters to future work. 

\section{Conclusion}
\label{sec:Conclusion}

In this work, we developed an extension of the Restriction Decoder introduced in Ref.~\cite{KD19} to the triangular color code. In the presence of boundaries, a new notion of connected components was introduced to deal with errors resulting in highlighted boundary vertices. We then showed how the Restriction Decoder can be adapted to include measurement errors. 

Next, we considered a fault-tolerant implementation of the triangular color code applicable to a full circuit level noise model. We showed how the triangular color code can be implemented on a lattice of degree three to reduce frequency collisions and cross talk errors relevant to superconducting qubit architectures. 
We emphasize that instead of the hexagonal lattice we could use the standard toric code architecture with the square lattice, where we do not use some connections between the qubits.
Further, we showed how the additional qubits necessary for a low degree implementation of the triangular color code can be used as flag qubits and provided a scheme which incorporates information from the flag qubits to preserve the full effective distance of the Restriction Decoder.  Performing a numerical simulation,
we estimated the storage threshold of the triangular color code against circuit-level depolarizing noise to be $0.2 \%$.
The threshold obtained is competitive with the surface code threshold, and thresholds for other low degree topological codes such as the ones considered in Ref.~\cite{CZYHC19}. However, the color code has the computational advantage that the full Clifford group can be implemented using only transversal operations. Hence due to the low degree connectivity of the hexagonal layout, transversal gate sets and competitive threshold, we believe the color code to be a promising candidate for fault-tolerant quantum computation.

For future work, it would be interesting to consider an implementation of the flag schemes presented in this work to other color code families, such as the 4.8.8 color code family and color codes with twist defects (see for instance \cite{TwistColorCode18}). Finding implementations of such codes on low degree graphs would also be an area of interest since such implementations could potentially be more suitable for superconducting qubit architectures. Further, of experimental relevance would be to study such codes under biased noise models such as in Refs.~\cite{CompassCodes,CompassV2}. Lastly, we believe a further numerical study for edge weight renormalizations, such as finding the optimal value of $\alpha$ (see the last paragraph of \cref{sec:NumRes}) would be of great interest. 

\section{Acknowledgements}

We thank Andrew Cross and Tomas Jochym-O'Connor for useful discussions.
We are grateful to Shilin Huang for pointing out to us an example of a weight-five error resulting in a logical failure for the distance $d = 13$ color code.
Further, we thank Andrew Cross for helping with parallelizing the C++ code when submitting jobs to the computing clusters.
We thank John Smolin for providing access to the computing clusters.
CC and GZ acknowledges partial support from Intelligence Advanced Research Projects Activity (IARPA) under contract W911NF-16-0114.
AK acknowledges funding provided by the Simons Foundation through the ``It from Qubit'' Collaboration.
Research at Perimeter Institute is supported in part by the Government of Canada through the Department of Innovation, Science and Economic Development Canada and by the Province of Ontario through the Ministry of Colleges and Universities.

\appendix

\section{A naive adaptation of the Restriction Decoder}
\label{sec_naive}

As we mentioned in \cref{subsec:AdaptDecTriangColor}, we can naively generalize the Restriction Decoder to the triangular color code as follows.
First, we pair up the restricted syndromes $\sigma_{RG}$ and $\sigma_{RB}$ within two restricted lattices $\dualx {RG}$ and $\dualx{RB}$.
The only difference from the original version of the Restriction Decoder is that now we allow vertices of the restricted syndromes to also be paired up with the boundary vertices.
We thus find for $C\in\{RG, RB\}$ a subset of edges $\rho_C \subseteq \face 1 {\dualx{C}}$, such that the $0$-boundary of $\rho_{C}$ is $\sigma_C\cup U$, where $U$ is some (possibly empty) subset of $\{v_R,v_G\}$ for $C=RG$ or $\{v_R,v_B\}$ for $C=RB$.
Then, we use the local lifting procedure \lift, which for every vertex $v\in\facex R 0 {\rho_{RG}\cup\rho_{RB}}$ finds a subset of faces $\tau_v\subseteq\star 2 v$, such that the $1$-boundary of $\tau_v$ locally matches $\rho_{RG}\cup\rho_{RB}$, i.e.,
$(\partial_2\tau_v)\rest v = (\rho_{RG}\cup\rho_{RB})\rest v$.
Note that whenever we pair up any of the vertices of the syndrome with the boundary vertex $v_R$, then we have to apply the lifting procedure \lift\ to $v_R$ (allowing $(\partial_2\tau_v)\rest v$ to differ from $(\rho_{RG}\cup\rho_{RB})\rest v$ by the edges $(v_R,v_G)$ and $(v_R,v_B)$).
Finally, we find the correction to be $\phi(\sigma) = \bigcup_{v\in\facex R 0 {\rho_{RG}\cup\rho_{RB}}} \tau_v$.

\begin{figure*}
\centering
(a)\includegraphics[width=0.37\textwidth]{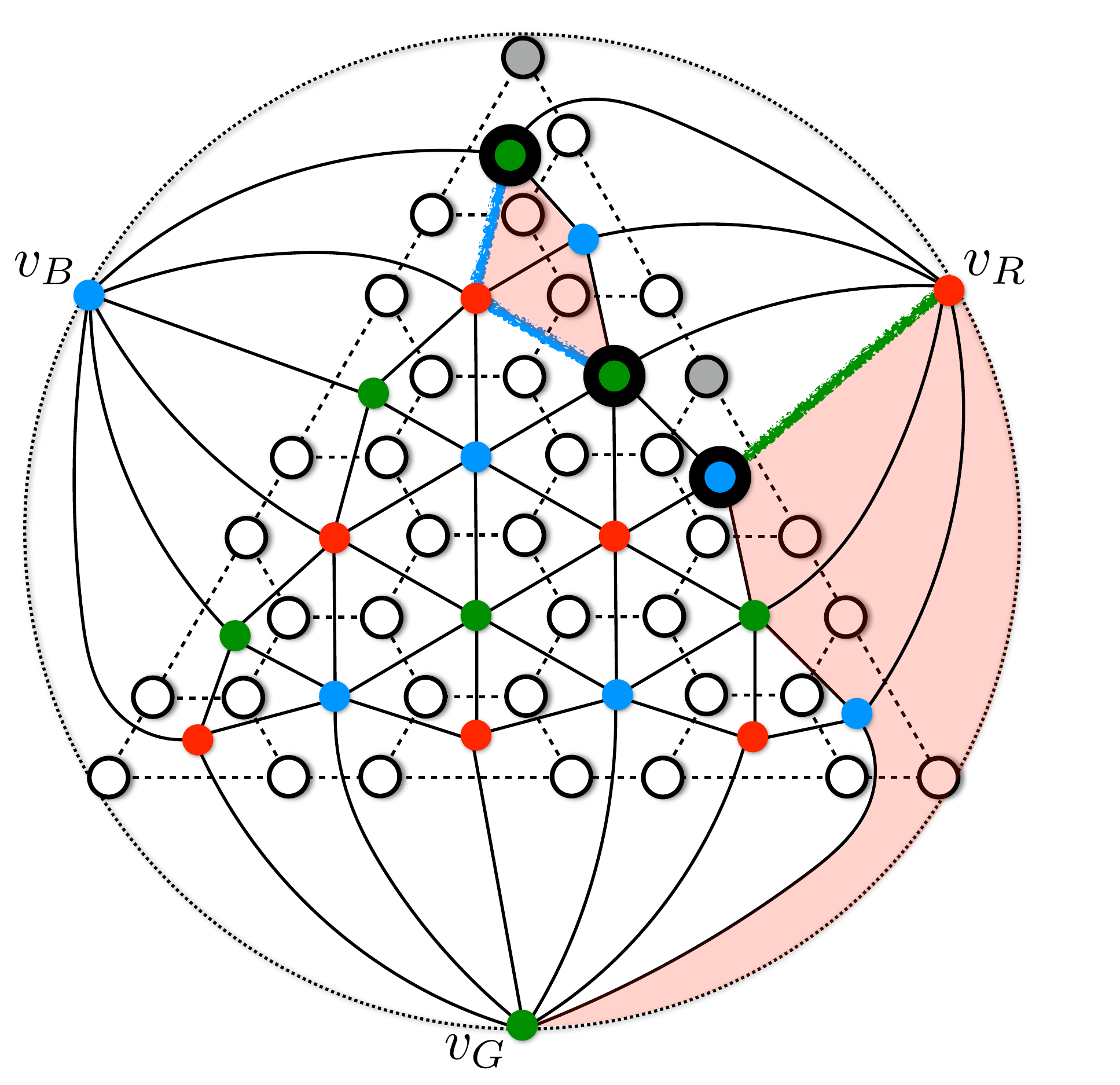}\hspace*{15mm}
(b)\includegraphics[width=0.37\textwidth]{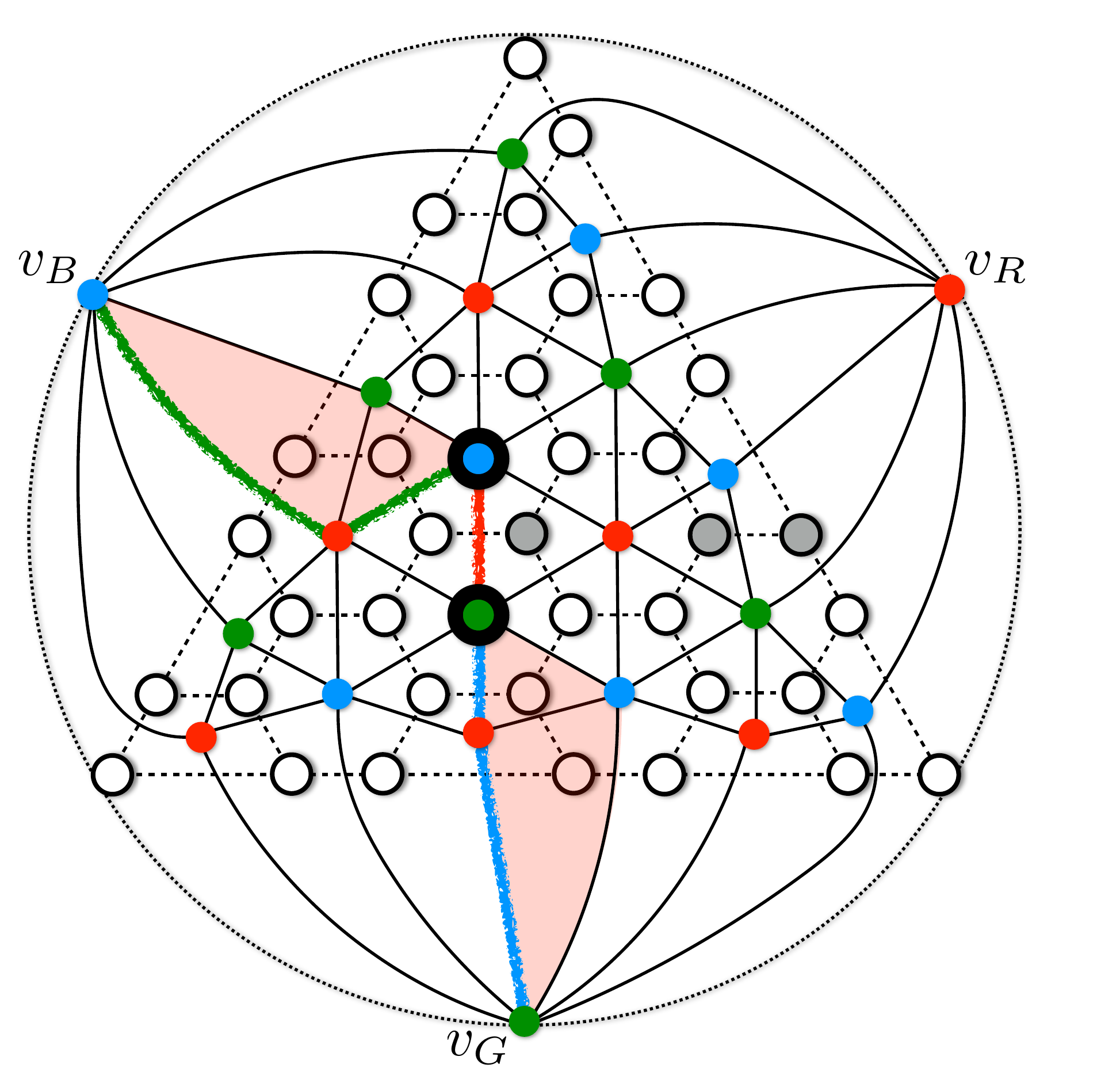}\\
(c)\includegraphics[width=0.37\textwidth]{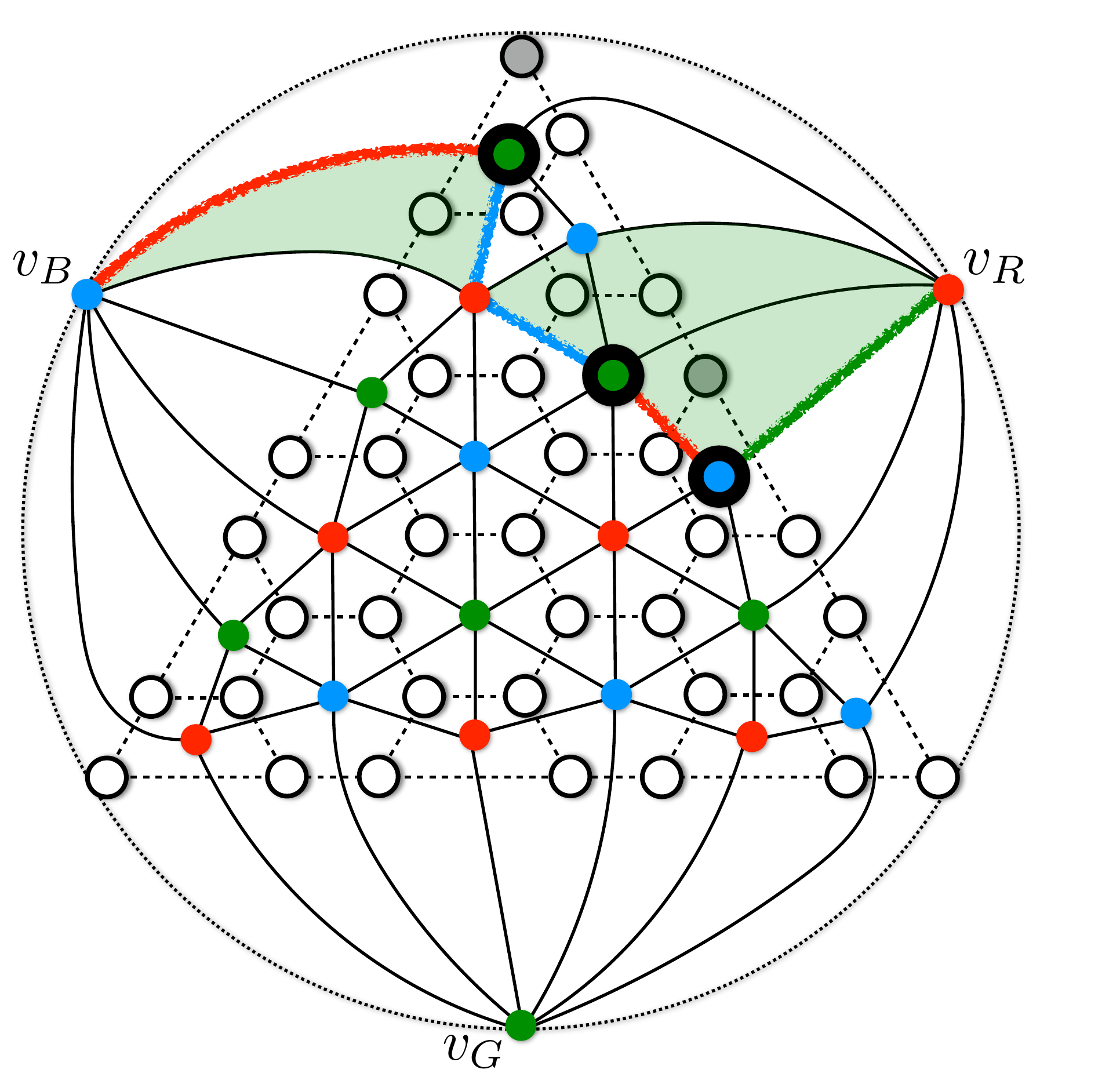}\\
\caption{
(a) A naive generalization of the Restriction Decoder to the triangular color code would pair up restricted syndromes $\sigma_{RG}$ and $\sigma_{RB}$ (marked vertices) within two restricted lattices $\dualx {RG}$ and $\dualx{RB}$, and apply the local lifting procedure to all vertices of color $R$, including the boundary vertex $v_R$.
Such a decoder is only guaranteed to correct errors of weight at most $\lfloor \frac{d-3}{4} \rfloor$.
On the other hand, we expect the effective distance for the adapted Restriction Decoder to scale as $\sim d/3$.
We illustrate examples of smallest weight errors (circles shaded in gray) leading to logical errors for (a) a naive generalization of the Restriction Decoder and (b) the adapted Restriction Decoder.
(c) The adapted Restriction Decoder successfully corrects the error, which in (a) led to a logical error when a naive generalization of the Restriction Decoder was used.
}
\label{fig_smallest_errors}
\end{figure*}

We illustrate a naive generalization of the Restriction Decoder in~\cref{fig_smallest_errors}.
The effective distance of that decoder is, roughly speaking, reduced by a factor of two compared to the color code distance.
In other words, a naive decoder is only guaranteed to correct errors of weight at most $\lfloor \frac{d-3}{4} \rfloor$.
On the other hand, we suspect that the adapted Restriction Decoder can correct all the errors of weight at most $\sim d/3$.
We illustrate examples of smallest weight errors leading to a logical error in ~\cref{fig_smallest_errors}.

\section{Edges and edge weight calculations for the space-time matching graphs $\match_{RG}$, $\match_{RB}$ and $\match_{GB}$}
\label{app:EdgeWeightCalc}

\begin{figure*}
	\centering
	(a)\includegraphics[width=0.37\textwidth]{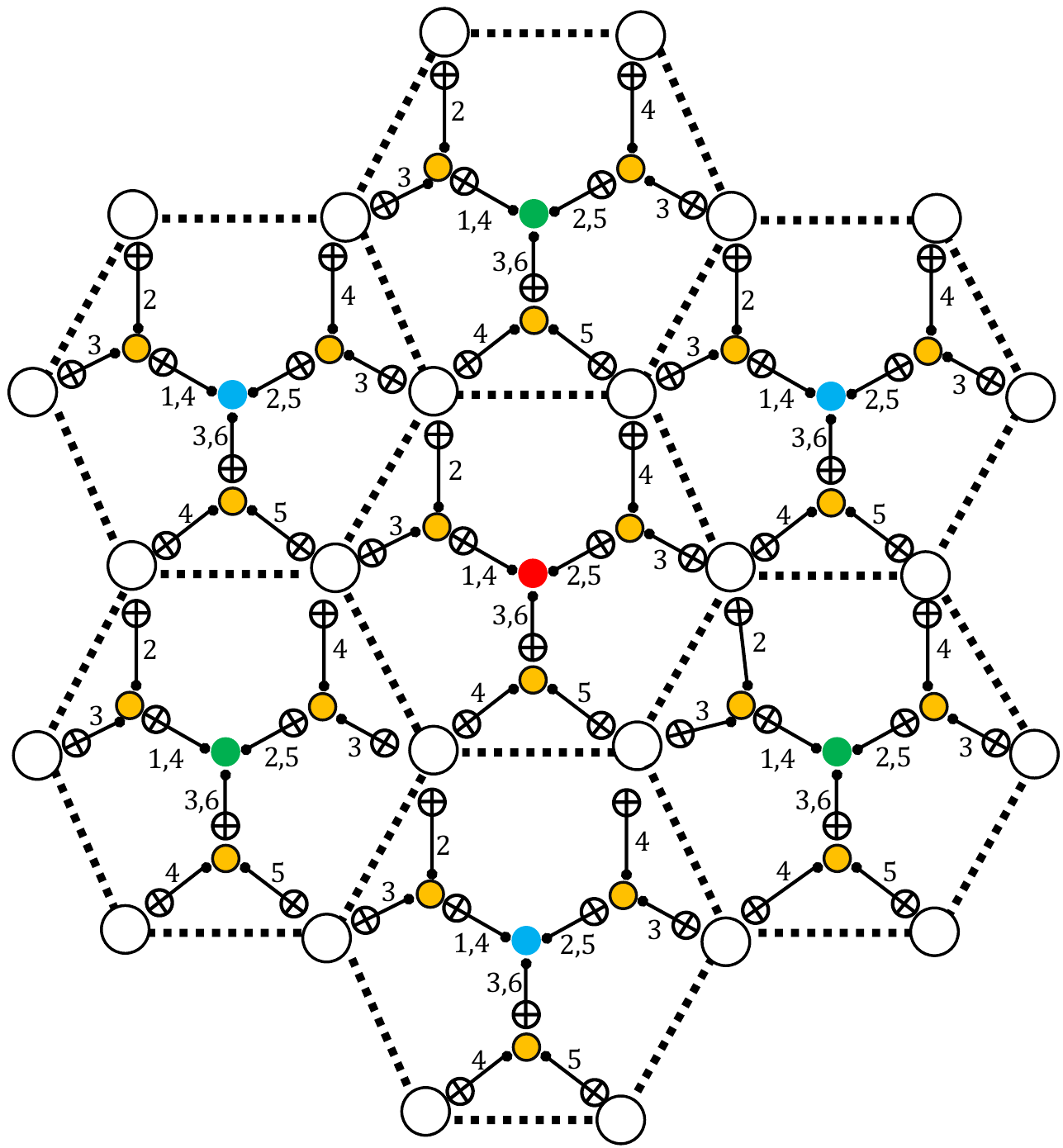}\hspace*{15mm}
        (b)\includegraphics[width=0.37\textwidth]{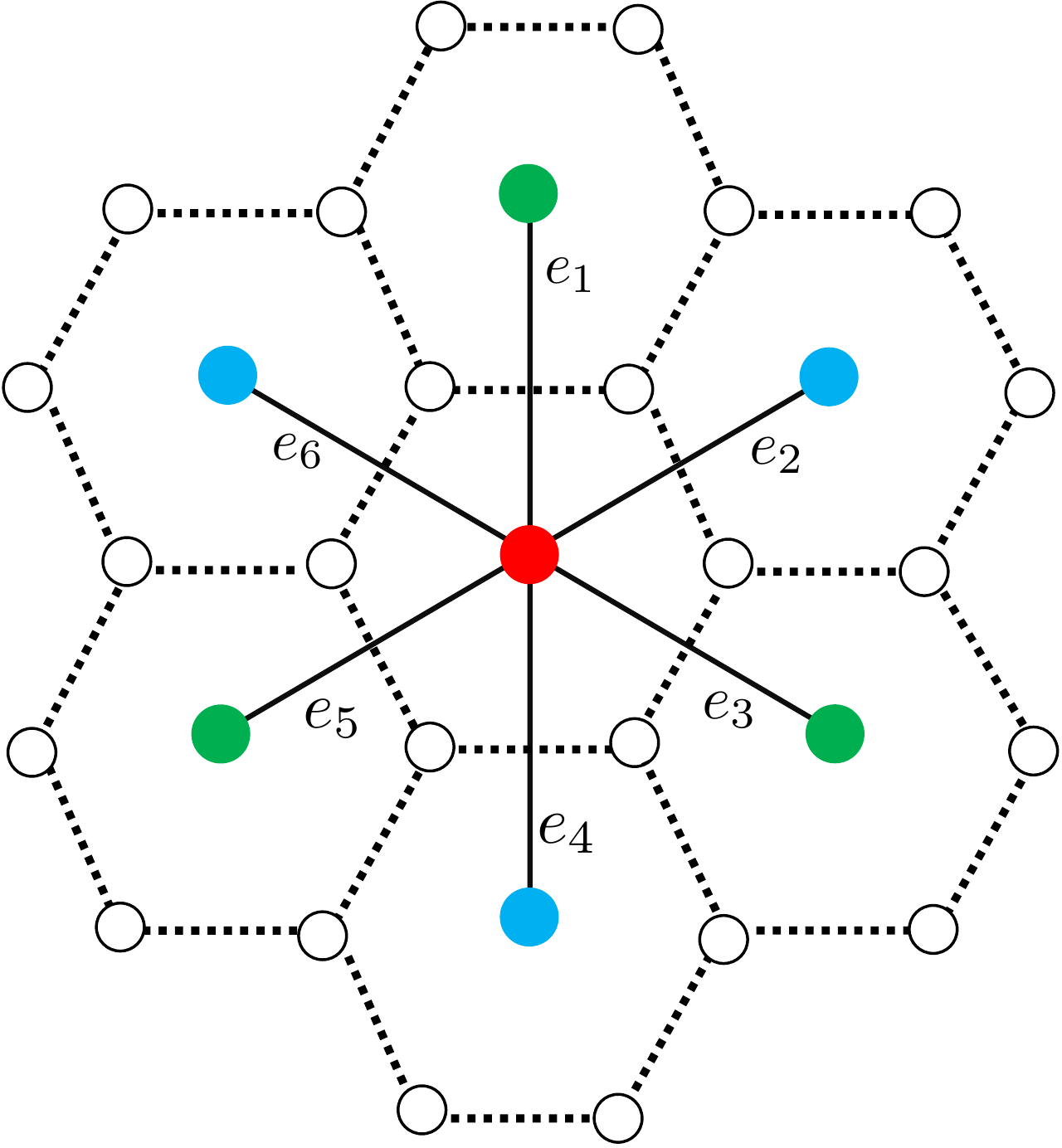}
	\caption{In (a), we show the connections between a red ancilla vertex, with blue and green vertices as would occur in the bulk of the triangular color code, with the layout shown in \cref{fig:FullScheduling}. In (b), we label the edges representing connections between the red vertex with the blue and green vertices found in the full RGB matching graph.}
	\label{fig:DiagEdgesCircuitCalc}
\end{figure*}

\begin{figure}
	\centering
	\includegraphics[width=0.17\textwidth]{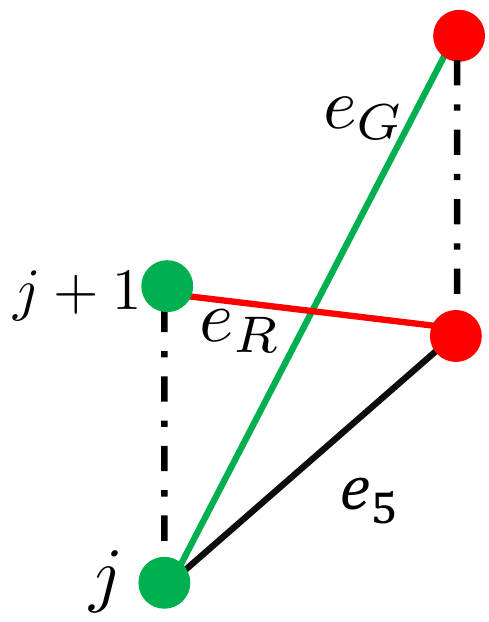}
	\caption{Diagonal edges $e_{R}$ and $e_{G}$ (shown in red and green) connecting red and green vertices between measurement rounds $j$ and $j+1$. If such an edge is chosen during MWPM, it will be projected to the edge $e_5$ when performing the flattening on edges of $\match_{RG}$.}
	\label{fig:DiagEdgeExample}
\end{figure}

In addition to the 2D edges of the lattice $\dual$, along with 3D vertical edges connecting a vertex in $\Delta_0 (\dual)$ of the same color in two different time steps to deal with measurement errors, there can also be correlated errors in both space and time arising from CNOT gate failures \cite{FowlerEdgeWeights}. To see this, consider the circuit in \cref{fig:DiagEdgesCircuitCalc} (a), which illustrates the connectivity between a red ancilla vertex in $\Delta_0 (\dual)$ (belonging to a face of $\mathcal{L}$ in the bulk), with blue and green vertices in $\Delta_0 (\dual)$. In \cref{fig:DiagEdgesCircuitCalc} (b), we label the edges between the red vertex with the green and blue vertices (as would be seen in $\dual$) as $e_1$ to $e_6$. Now, consider the CNOT gates connecting the red and green vertices to the data qubits (represented by yellow circles), along the edge $e_5$. In particular, we focus on the CNOT gates between the flag qubits (represented by white circles) and the data qubits. 

\begin{figure*}
	\centering
	(a)\includegraphics[width=0.6\textwidth]{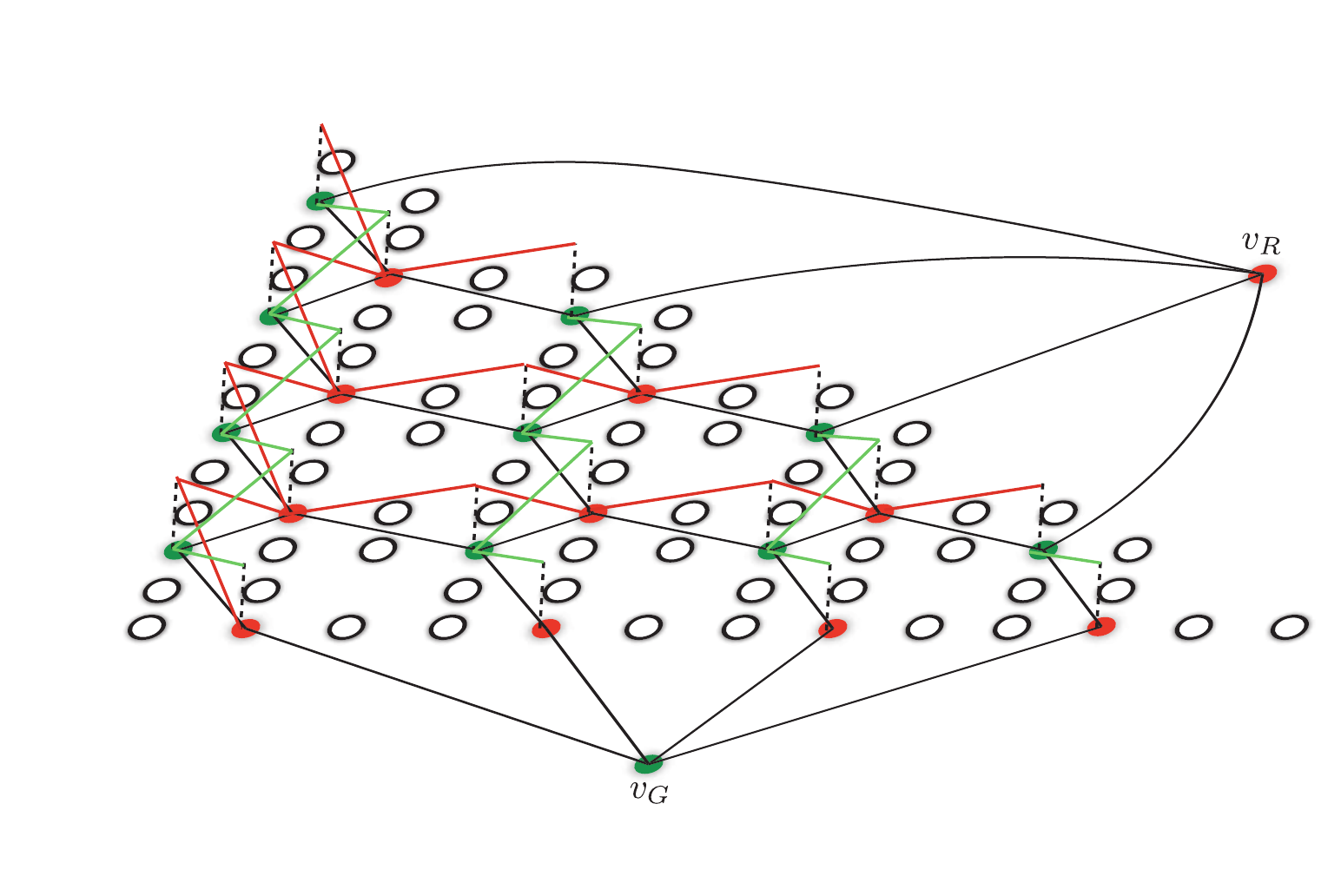}\\
        (b)\includegraphics[width=0.6\textwidth]{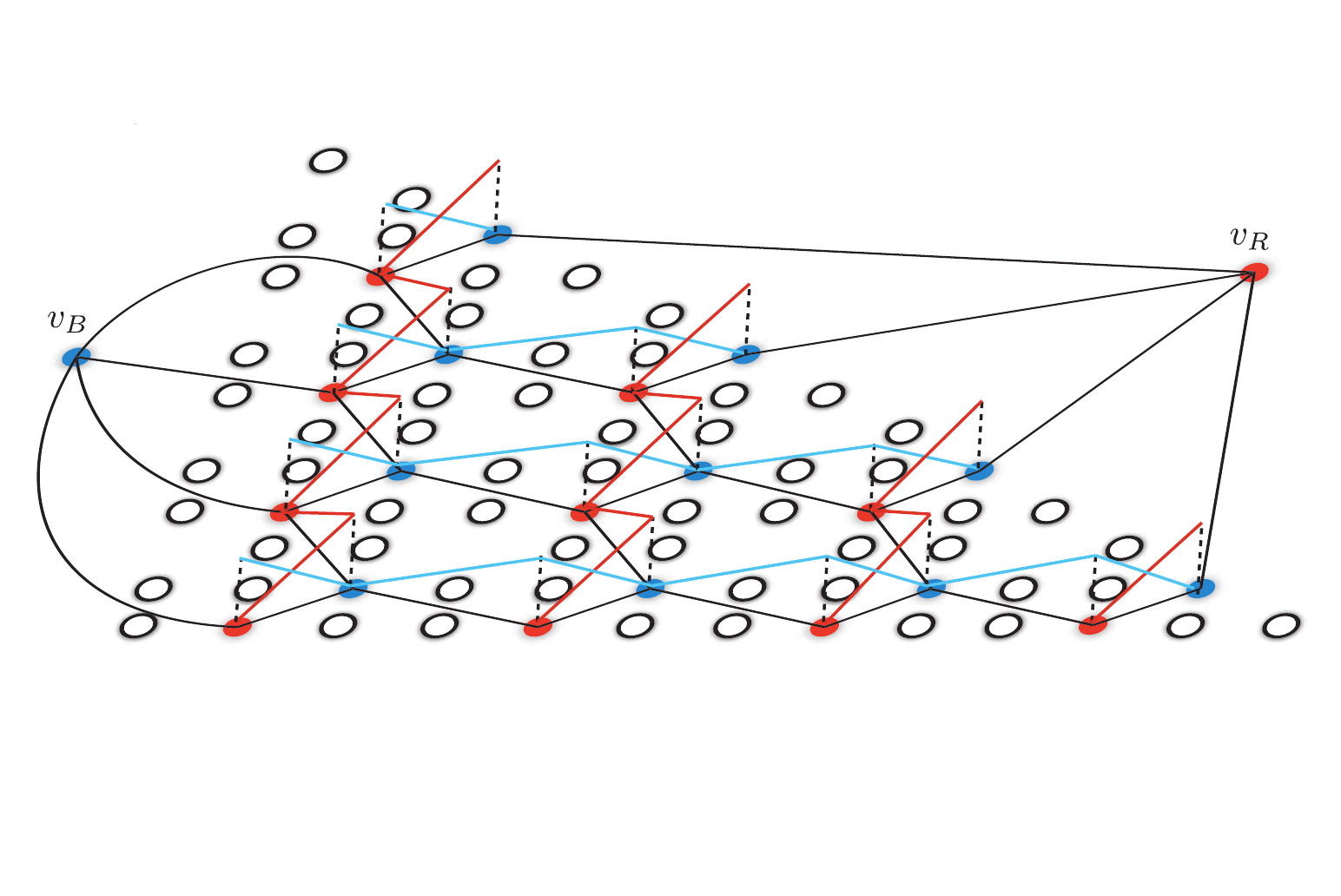}\\
        (c)\includegraphics[width=0.6\textwidth]{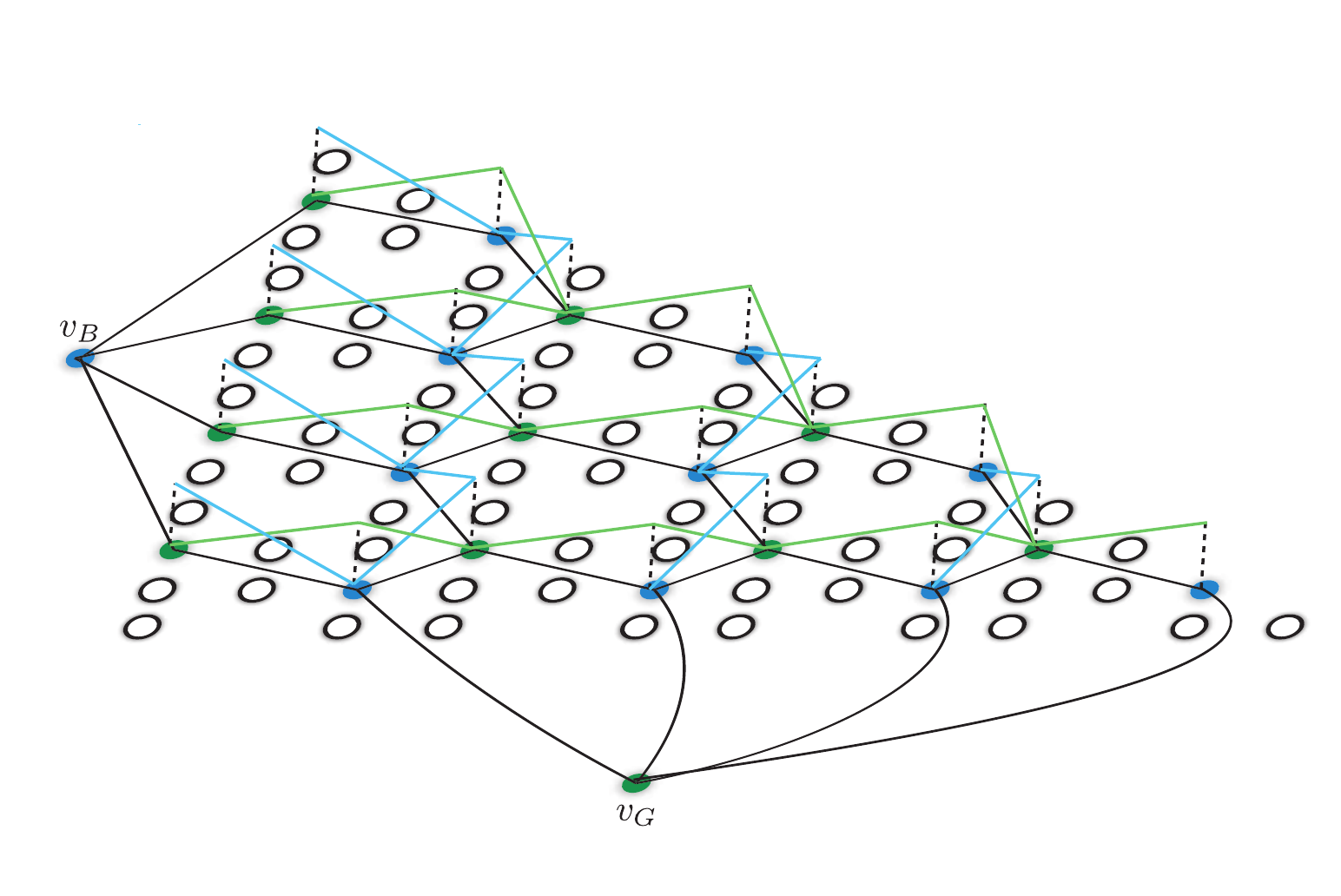}
	\caption{Illustration of all 3D diagonal edges connecting two different 2D lattices for the matching graphs (a) $\match_{RG}$ , (b) $\match_{RB}$ (c) $\match_{GB}$. Note that the flag edges are not shown, and vertical edges are represented by dashed lines.}
	\label{fig:3DEdgePlot}
\end{figure*}

Let $C_{t}^{l}$ correspond to a CNOT gate belonging to a face of $\mathcal{L}$ with a vertex in $\Delta_0 (\dual)$ of color $l$, applied during time step $t$ for a given round of syndrome measurements. Now suppose that during the $j$'th syndrome measurement round, the CNOT $C_{4}^{G}$ fails and introduces an error from the set $\{ ZZ, ZY, YZ, YY \}$. Propagating such errors through the stabilizer measurement circuits of \cref{fig:StabCircMeas} (b), one can show that such a fault introduces a $Z$ error on the data qubit $q_j$ shown in \cref{fig:DiagEdgesCircuitCalc} (a). However, given the time step at which the error occurs (the fourth time step), we find that only the green ancilla vertex is highlighted. If a $Z$ error on the data qubit $q_j$ had instead occurred during the first time step of the syndrome measurement round $j$, then both green and red ancillas would have been highlighted (assuming no other errors were introduced). Now during the next syndrome measurement round $j+1$ (again, assuming no other errors are introduced), both red and green ancillas would be highlighted. Similarly, if the CNOT gate $C_3^{R}$ failed and introduced an error from the set $\{ IZ, XZ, IY, XY \}$, the same pattern in highlighted ancillas would be observed.

Now, if both red and green ancillas were highlighted in the same syndrome measurement round (say during the round $j$, caused by a $Z$ data qubit error on qubit $q_j$ at the first time step), the edge $e_5$ would be chosen when performing MWPM on the matching graph $\match_{RG}$.  However, for CNOT failures mentioned above, since the red vertex is highlighted during round $j+1$, whereas the green vertex is highlighted in both rounds $j$ and $j+1$, one can add the green edge $e_{G}$ shown in \cref{fig:DiagEdgeExample} to the matching graph $\match_{RG}$. Since the Restriction Decoder considers changes in measurement outcomes of a given vertex between consecutive syndrome measurement rounds, the set of highlighted vertices $W$ (see \cref{Subsec:IncMeasErrs}) will contain the green vertex for round $j$ with the red vertex for round $j+1$. Thus the shortest path connecting both vertices is obtained by choosing the edge $e_{G}$. If such an edge was chosen during MWPM, $e_{G}$ would be projected onto the edge $e_{5}$ when applying the flattening map $g$ (see \cref{eq:Flattg}). Lastly, it can be shown that the red edge $e_{R}$ in \cref{fig:DiagEdgeExample} would be chosen if the CNOT $C_3^{G}$ failed and introduced an error from the set $\{ IZ, XZ, IY, XY \}$, or the CNOT $C_4^{G}$ failed introducing errors from the set $\{ ZZ, ZY, YZ, YY \}$. 3D edges as the ones shown in \cref{fig:DiagEdgeExample} are referred to as 3D diagonal edges since they are due to errors arising from CNOT gates resulting in different highlighted vertices in $W$ between two consecutive syndrome measurement rounds. Consequently, such faults result in highlighted edges between vertices belonging to different locations in the 2D lattice $\dual$ (after performing the flattening on all edges belonging to the matching graphs $\match_{RG}$, $\match_{RB}$ and $\match_{GB}$). Measurement errors result in 3D vertical edges connecting the same vertex in two different syndrome measurement rounds (see for instance \cite{FMMC12}).

The set of all 3D diagonal edges associated with the matching graphs $\match_{RG}$, $\match_{RB}$ and $\match_{GB}$ can be found in \cref{fig:3DEdgePlot}. Such edges are obtained by performing a similar analysis to the one performed above leading to the edges shown in \cref{fig:DiagEdgeExample}. In particular, we considered all single fault events arising from CNOT failures leading to edges which (after applying the flattening map $g$) are projected on the 2D edges $e_1$ to $e_6$ in the bulk (see \cref{fig:DiagEdgesCircuitCalc} (b)). In addition, we also considered the CNOT scheduling at all boundary locations. 

\begin{figure*}
	\centering
	(a)\includegraphics[width=0.6\textwidth]{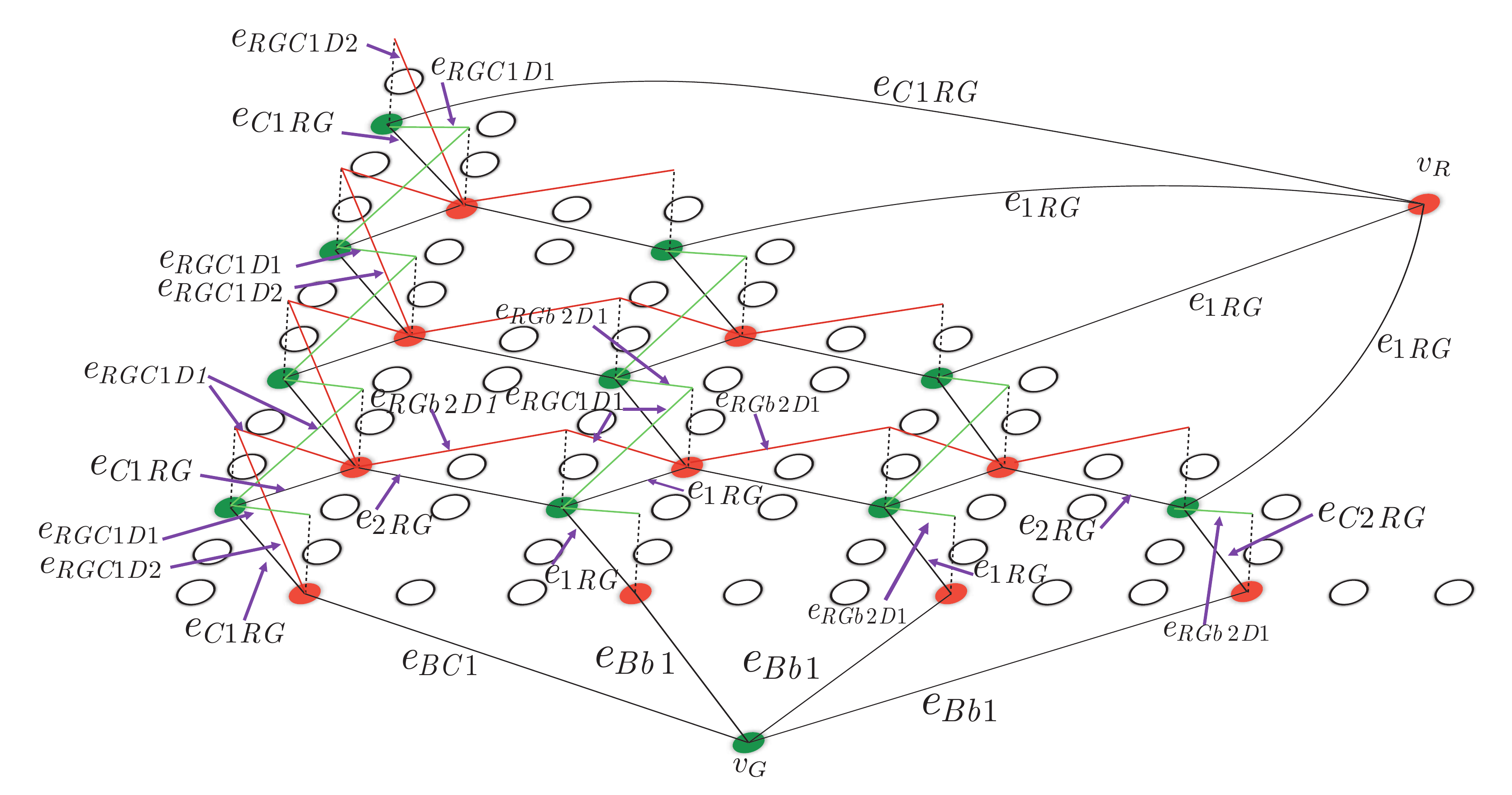}
	(b)\includegraphics[width=0.6\textwidth]{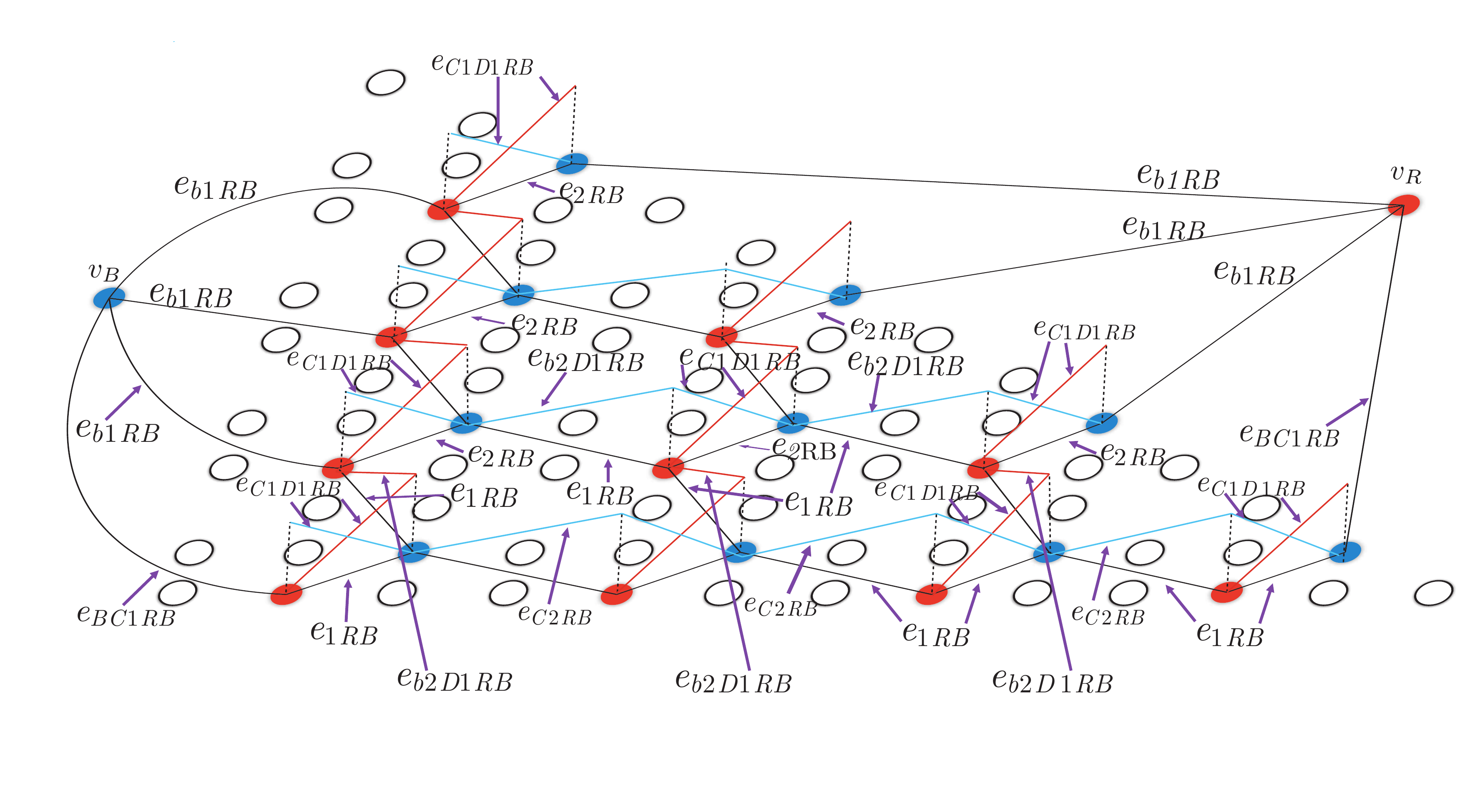}
	(c)\includegraphics[width=0.6\textwidth]{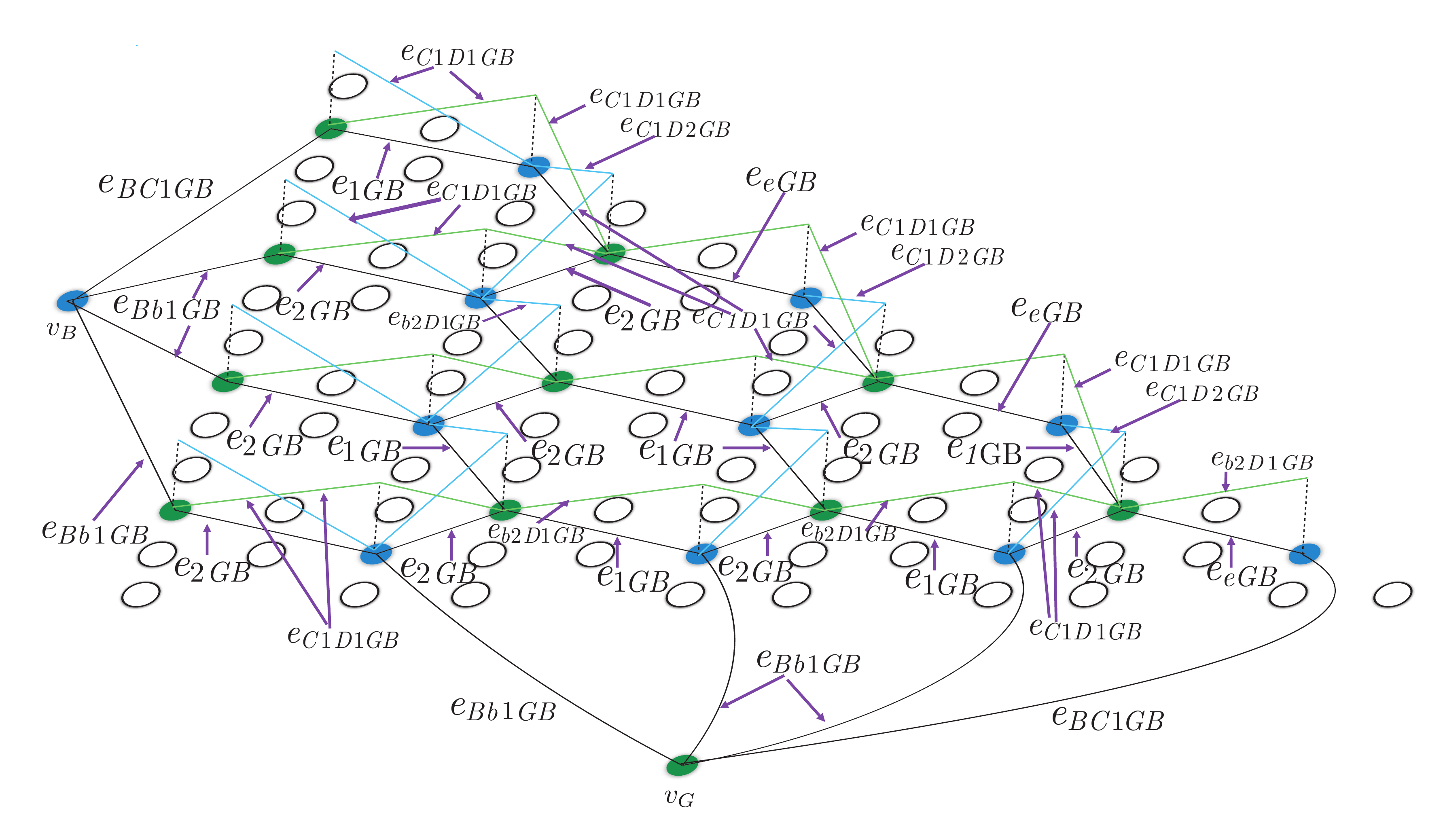}
	\caption{Edge labels for the (a) $\match_{RG}$,  (b) $\match_{RB}$ graph and (c) $\match_{GB}$ graphs. Edges with identical labels have the same weights. Edges in the bulk with different weights are only labelled once since such patterns are repeated throughout the bulk. }
	\label{fig:3DedgeLabels}
\end{figure*}

\begin{table}
	\begin{centering}
		\begin{tabular}{|c|}
			\hline
$p_{e_{1RG}} = p_{e_{1RB}} = p_{e_{1GB}} =  \frac{44}{15}p + \mathcal{O}(p^2)$  \\ \hline
$p_{e_{2RG}} = p_{e_{2RB}} = p_{e_{2GB}} = \frac{52}{15}p + \mathcal{O}(p^2)$  \\ \hline
$p_{e_{C1RG}} = \frac{12}{5}p + \mathcal{O}(p^2)$  \\ \hline
$p_{e_{mw6}} = \frac{76}{15}p + \mathcal{O}(p^2)$  \\ \hline
$p_{e_{mw4}} = \frac{64}{15}p + \mathcal{O}(p^2)$  \\ \hline
$p_{e_{RGC1D1}} = \frac{8}{15}p + \mathcal{O}(p^2)$  \\ \hline
$p_{e_{RGC1D2}} = \frac{16}{15}p + \mathcal{O}(p^2)$  \\ \hline
$p_{e_{RGb2D1}} = \frac{8}{5}p + \mathcal{O}(p^2)$  \\ \hline
$p_{e_{BC1RB}} = \frac{44}{15}p + \mathcal{O}(p^2)$  \\ \hline
$p_{e_{b1RB}} = \frac{52}{15}p + \mathcal{O}(p^2)$  \\ \hline
$p_{e_{C1D1RB}} = \frac{8}{15}p + \mathcal{O}(p^2)$  \\ \hline
$p_{e_{C2RB}} = \frac{16}{15}p + \mathcal{O}(p^2)$  \\ \hline
$p_{e_{b2D1RB}} = \frac{8}{5}p + \mathcal{O}(p^2)$  \\ \hline
$p_{e_{eGB}} = \frac{12}{5}p + \mathcal{O}(p^2)$  \\ \hline
$p_{e_{BC1GB}} = \frac{44}{15}p + \mathcal{O}(p^2)$  \\ \hline
$p_{e_{Bb1GB}} = \frac{52}{15}p + \mathcal{O}(p^2)$  \\ \hline
$p_{e_{C1D1GB}} = \frac{8}{15}p + \mathcal{O}(p^2)$  \\ \hline
$p_{e_{b2D1GB}} = \frac{8}{5}p + \mathcal{O}(p^2)$  \\ \hline
$p_{e_{C1D2GB}} = \frac{16}{15}p + \mathcal{O}(p^2)$  \\ \hline
		\end{tabular}
		\par\end{centering}		
	\caption{\label{Tab:EdgePolyList} List of edge weight polynomials for the edges labelled in \cref{fig:3DedgeLabels}. Since the polynomials were computed by considering leading order events resulting in a highlighted edge, only leading order terms in $p$ are provided. The polynomials $p_{mw6}$ and $p_{mw4}$ apply to vertical edges arising from measurement errors during weight-four and weight-six stabilizer measurements, and are the same for all graphs $\match_{RG}$, $\match_{RB}$ and $\match_{GB}$.}
\end{table}

Next, we show how to obtain the edge weight for $e_{G}$ in \cref{fig:DiagEdgeExample}. From the noise model described in \cref{subsubEdgeRen}, each two-qubit Pauli operator occurs with probability $p/15$. To leading order in $p$, a green edge will occur if the CNOT $C_{4}^{G}$ fails introducing an error from the set $\{ ZZ, ZY, YZ, YY \}$ (which has a total probability of $4p/15$) and no failure occurs for $C_3^{R}$, or $C_3^{R}$ fails introducing an error from the set $\{ IZ, XZ, IY, XY \}$ and no failure occurs for $C_{4}^{G}$. Summing the probabilities for both cases, we have (to leading order in $p$) that the total probability of obtaining a green highlighted edge in \cref{fig:DiagEdgeExample} is
\begin{align}
p_{e_{G}} = \frac{8p}{15}\left(1-\frac{4p}{15}\right),
\label{eq:EgePoly}
\end{align}
and thus the edge weight is given by $w_{e_{G}}=-\log{p_{e_{G}}}$. In what follows, polynomials $p_{e}$ for an edge $e$ (as in \cref{eq:EgePoly}) will be referred to as edge weight polynomials. 

In \cref{fig:3DedgeLabels}, the edges for the graphs  $\match_{RG}$, $\match_{RB}$ and $\match_{GB}$ are labeled in such a way that edges with the same label have identical edge weights. Further, in \cref{Tab:EdgePolyList}, we provide the edge weight polynomials $p_{e}$ for the edges $e$ shown in \cref{fig:3DedgeLabels}. Since the edge weights were computed by considering only leading order events (i.e. single fault events which result in such edges to be the shortest path between highlighted vertices in $\match_{C}$), only the terms to leading order in $p$ are provided.

\section{Proof of Theorem~\ref{thm:1flag_suff}}
\label{app:more_thm}
In this appendix provide the proof of Theorem~\ref{thm:1flag_suff} which we restate here for convenience:

\textbf{Theorem III.1}
The triangular color code can be decoded with full distance if stabilizers are measured with 1-flag circuits.

\begin{proof}
To prove the theorem, it is enough to show that, within a single round of stabilizer measurements, any set of $d-1$ or fewer faults does not cause a (nontrivial) logical error to be placed on the data without any flag qubits flagging. This is enough because it implies that two different sets of $\lfloor(d-1)/2\rfloor$ or fewer faults can be distinguished either by their flags or by future rounds of stabilizer measurements. Therefore, in principle, there is an algorithm to decode with full distance.

There is a simplification to the problem coming from the form of the syndrome extraction circuits. Notice that if the circuits measuring $Z$- or $X$-type stabilizers propagate errors from ancilla to data qubits, then the resulting errors on the data are $Z$ or $X$ errors, respectively. If it takes $f_0$ faults to place a nontrivial logical operator $P$ without flag qubits flagging, then there is a way to use $f\le f_0$ faults to place the logical operator $Z(P)$ and $f'\le f_0$ faults to place logical operator $X(P)$, where $Z(P)$ is an $Z$-type Pauli string with Pauli $Z$ wherever $P$ has $Z$ or $Y$ and with identity elsewhere and $X(P)=PZ(P)$. As $P$ is nontrivial, at least one of $Z(P)$ and $X(P)$ is also nontrivial while being entirely $Z$- or $X$-type. Using the code symmetry, assume $Z(P)$ is nontrivial. In the remainder, we restrict attention to errors of purely $Z$-type and prove a lower bound $f_0\ge f\ge d$.

Given a logical Pauli operator $P$ of $Z$-type, we can associate parts of its support to each face. During a single round of syndrome extraction, imagine that faults in the circuit measuring the stabilizer on face $i$ result in $Z$ errors on qubits in a set $A_i$ (of course $A_i$ must be a subset of the qubits in the face). Then, if there are $s$ faces and $\triangle$ is the symmetric difference operation on sets, we call the collection of sets $\{A_i\}$ an \emph{over-partition} of $\supp{P}$ because $\supp{P}=A_1\triangle A_2\triangle\dots\triangle A_s$. If $A_i\cap A_j=\emptyset$ for all $i\neq j$, then $\{A_i\}$ is actually a partition, $\supp{P}=A_1\cup A_2\cup\dots\cup A_s$. Let $a_x=|\{i:|A_i|=x\}|$ for $x=0,1,\dots,6$ count the number of sets of size $x$ (we always use lower case for counts corresponding to upper case sets). Of course, $\sum_{x=0}^6a_x=s$ but notice also that $|\supp{P}|\le\sum_{x=1}^6xa_x$ for any over-partition with equality if and only if $\{A_i\}$ is a partition. A given logical operator may have multiple over-partitions and multiple partitions. There is always a partition $\{A'_i\}$ for every over-partition $\{A_i\}$ such that $A'_i\subseteq A_i$ (formed, for instance, by repeatedly finding a qubit $q$, if it exists, that appears in two sets $A_i$ and $A_j$ and removing it from both). We call $\{A'_i\}$ a \emph{sub-partition} of $\{A_i\}$.

We use two facts for this proof -- (1) by definition (see Ref.~\cite{CB17}), if a 1-flag circuit for syndrome extraction results in two or more errors on the data, either more than one fault has occurred in the circuit or a flag has flagged and (2) $|\text{supp}(P)|\ge d+a'_3+2a'_4+4a'_5+6a'_6$ for any partition $\{A'_i\}$ (not for an over-partition). This second fact depends on structure of the triangular color code, and we provide the proof below.

From these two facts the conclusion follows. By fact (1), it takes at least two faults to create errors on all qubits in $A_i$ if $|A_i|\ge2$. This implies a total of $f\ge a_1+2\sum_{x=2}^6a_x=2s-a_1-2a_0$ faults. A sub-partition $\{A'_i\}$ of $\{A_i\}$ satisfies $2s-a_1-2a_0\ge 2s-a'_1-2a'_0$. We want to show $f\ge d$, and by the prior discussion it is enough to show $2s-a'_1-2a'_0\ge d$. But this is easily done with fact (2) implying $\sum_{x=1}^6xa'_x=|\supp{P}|\ge d+a'_3+2a'_4+4a'_5+6a'_6$ whose rearrangement shows $d\le a'_1+2a'_2+2a'_3+2a'_4+a'_5\le f$.
\end{proof}

We now fill the last step in the proof of \cref{thm:1flag_suff}. We termed this ``fact (2)" in that proof, and it says that any partition $\{A_i\}$ of a logical operator $P$ satisfies $|\text{supp}(P)|\ge d+a_3+2a_4+4a_5+6a_6$. Refer to the theorem for notation.

Since $\{A_i\}$ is a partition for a logical operator $P$, replacing $A_i$ with $\overline{A}_i$ (the complement of $A_i$ within the support of face $i$) is an over-partition for another logical operator, namely $P$ times a stabilizer on face $i$. We therefore can choose $B_i=A_i$ if $|A_i|\le3$ and $B_i=\overline{A}_i$ if $|A_i|\ge4$ to get an over-partition for a logical operator $Q$. Then $b_4=b_5=b_6=0$, $b_0\ge a_0+a_6$, $b_1=a_1+a_5$, $b_2\le a_2+a_4$ and $b_3=a_3$. If all faces were hexagonal, we would have equality in all these relations between $b_x$ and $a_x$, but the presence of square faces means the inequalities are correct. We have $d\le|\supp{Q}|\le b_1+2b_2+3b_3\le |\supp{P}|-2a_4-4a_5-6a_6:=N$. Note that this takes care of fact (2) when $a_3=b_3=0$.

We now prove the following Lemma.
\begin{lemma}
If $Q$ is a logical operator with over-partition $\{B_i\}$ such that $|B_i|\le3$ for all $i$, $b_3>0$, and $|\supp{Q}|\le b_1+2b_2+3b_3\le N$, then there is another logical operator $Q'$ with over-partition $\{B'_i\}$ such that $|B'_i|\le 3$ for all $i$, $b'_3<b_3$, and $|\supp{Q'}|\le b'_1+2b'_2+3b'_3\le N-(b_3-b'_3)$.
\end{lemma}
\noindent Note that the repeated application of the lemma to $Q$ and $N$ defined prior immediately implies that there is a logical operator $R$ with over-partition $\{G_i\}$ such that $g_3=g_4=g_5=g_6=0$ and $d\le|\supp{R}|\le N-a_3$. Plugging $N$ into this last inequality completes the proof of fact (2).

This lemma's proof proceeds in two steps. Step (a) is to create a sub-partition $\{C_i\}$ for $Q$, so $|\supp{Q}|=c_1+2c_2+3c_3\le(b_1+2b_2+3b_3)-(b_3-c_3)$. The inequality holds because $b_3-c_3$ sets $B_i$ with size three will become sets $C_i$ with size two or less during the sub-partition algorithm. It may be that $c_3=b_3$, so we are not done.

Step (b) begins by finding a face $i$ such that $|C_i|=3$ (if none exists we are done). Since $Q$ is logical, it must be that it commutes with the stabilizer on face $i$ and so has even overlap with the face’s support. This implies another qubit $q\in\supp{Q}$ but $q\not\in C_i$. So $q\in C_j$ for some $j\neq i$. Move $q$ from $C_j$ to $C_i$ and take the complement of set $i$. This defines a new over-partition $\{B'_i\}$ for another logical operator $Q'$. Note further that
\begin{align}
|\supp{Q'}|&\le b'_1+2b'_2+3b'_3\\
&=\bigg\{\begin{array}{ll}|\supp{Q}|-2,&\text{face $i$ is hexagon}\\|\supp{Q}|-4,&\text{face $i$ is square}\end{array}\\
&\le|\supp{Q}|-2
\end{align}
and that $b_3\ge c_3>b'_3\ge c_3-2$. So $|\supp{Q'}|\le b'_1+2b'_2+3b'_3\le|\supp{Q}|-(c_3-b'_3)\le N-(b_3-b'_3)$.

\clearpage

\bibliographystyle{unsrtnat} 
\bibliography{bibtex_chamberland}

\begin{thebibliography}{60}
\bibitem[TC_()]{TC_threshold}
The threshold of $p_{\text{th}} = 0.7\%$ was obtained from multiple independent
  simulations performed at IBM using the noise model described in the text and
  optimized edge weights. Note that in \cite{FowlerEdgeWeights}, a threshold of
  over $1\%$ was obtained, but we could not reproduce such results for the
  noise model considered in this work.
\end{thebibliography}

\end{document}